\documentclass[noacm]{acmart}

\usepackage{booktabs} 
\usepackage{ccicons}  
\usepackage{subfig}
\graphicspath{{figures/}}
\usepackage[english]{babel}
\usepackage[autostyle=true,english=american]{csquotes}
\usepackage{gensymb}
\usepackage{balance}
\usepackage{wrapfig}
\usepackage{physics}
\usepackage{hyperref}

\usepackage{acronym}
\newacro{AI}[AI]{Artificial Intelligence}
\newacro{UI}[UI]{user interface}
\newacro{GUI}[GUI]{graphical user interface}
\newacro{TLX}[TLX]{NASA-Task Load Index}
\newacro{RTLX}[Raw-TLX]{NASA Raw-Task Load Index}
\newacro{ER}[ER]{error rate}

\newacro{TCT}[TCT]{task completion time}
\newacro{HCI}[HCI]{Human-Computer Interaction}
\newacro{UX}[UX]{user experience}
\newacro{HFE}[HFE]{Human Factors and Ergonomics}
\newacro{cuDNN}[cuDNN]{CUDA Deep Neural Network library}
\newacro{RMSE}[RMSE]{root mean squared error}
\newacro{HMD}[HMD]{Head-Mounted Display}
\newacro{RF}[RF]{Random Forest}
\newacro{GP}[GP]{Gaussian process, long-plural = Gaussian processes}
\newacro{KNN}[\textit{k}NN]{\textit{k}-nearest neighbor}
\newacro{NN}[NN]{Neural Network}
\newacro{DNN}[DNN]{ Deep Neural Network}
\newacro{CNN}[CNN]{Convolutional Neural Network}
\newacroplural{CNN}[CNNs]{Convolutional Neural Networks} 
\newacro{FCL}[FCL]{fully connected layer}
\newacro{BoD}[BoD]{Back-of-Device}
\newacro{FOV}[FoV]{field of view}
\newacro{RW}[RW]{Real World}
\newacro{IFRC}[IFRC]{index finger ray cast}
\newacro{FRC}[FRC]{forearm ray cast}
\newacro{EFRC}[EFRC]{eye-finger ray cast}
\newacro{HRC}[HRC]{Human-Robot Collaboration}
\newacro{HRI}[HRI]{Human-Robot Interaction}
\newacro{6DOF}[6DOF]{six-degree-of-freedom}
\newacro{LOOCV}[LOOCV]{leave-one-out cross-validation}
\newacro{CV}[CV]{cross-validation}
\newacro{RM}[RM]{repeated measure}
\newacro{ANOVA}[ANOVA]{Analysis of Variance}
\newacro{RMANOVA}[RM-ANOVA]{rRepeated Measures Analysis of Variance}
\newacro{AGATe}[AGATe]{AGreement Analysis Toolkit}
\newacro{GHoST}[GHoST]{Gesture Heatmap Toolkit Gesture Heatmaps Toolkit}
\newacro{GREAT}[GREAT]{Gesture Relative Accuracy Toolkit}
\newacro{GRT}[GRT]{Gesture Recognition Toolkit}
\newacro{DTW}[DTW]{Dynamic Time Warping}
\newacro{LHRD}[LHRD]{large high resolution display}
\newacro{GEQ}[GEQ]{Game Experience Questionnaire}
\newacro{SPGQ}[SPGQ]{Social Presence Gaming Questionnaire}
\newacro{JND}[JND]{just-noticeable difference}
\newacro{SUS}[SUS]{system usability scale}
\newacro{CSCW}[CSCW]{computer-supported cooperative work}
\newacro{CAD}[CAD]{computer-aided design}
\newacro{MR}[MR]{Mixed Reality}
\newacro{CVE}[CVE]{Collaborative Virtual Environment}
\newacro{AR}[AR]{Augmented Reality}
\newacro{AV}[AV]{Augmented Virtuality}
\newacro{VR}[VR]{Virtual Reality}
\newacro{PRISMA}[PRISMA]{Preferred Reporting Items for Systematic Reviews}
\newacro{PRISMA-Scope}[PRISMA-ScR]{Meta-Analyses Extension for Scoping Reviews}
\newacro{TF-IDF}[TF-IDF]{Term Frequency-Inverse Document Frequency}
\newacro{TF}[TF]{Term Frequency}
\newacro{AVs}[AVs]{Automated Vehicles}
\newacro{eHMIs}[eHMIs]{external Human-machine interfaces}
\newacro{SAR}[SAR]{Spatial Augmented Reality}
\newacro{MAR}[MAR]{Mobile Augmented Reality}
\newacro{IFR}[IFR]{International Federation of Robotics}
\newacro{ADLs}[ADLs]{Activities of Daily Living}
\newacro{LED}[LED]{Light-Emitting Diode}
\newacro{DoF}[DoF]{Degree-of-Freedom}
\newacroplural{DoF}[DoFs]{Degrees-of-Freedom}  
\newacro{HHC}[HHC]{Human-Human Collaboration}
\newacro{IDF}[IDF]{Inverse Document Frequency}
\newacro{QUEAD}[QUEAD]{Questionnaire for the Evaluation of Physical Assistive Devices}
\newacro{TiA}[TiA]{Trust in Automation Questionnaire}
\newacro{TOR}[TOR]{Take-Over-Request}
\newacro{ADMC}[ADMC]{Adaptive DoF Mapping Control}
\newacro{ML}[ML]{Machine Learning}
\newacro{IR}[IR]{infrared}
\newacro{IK}[IK]{inverse kinematics}
\newacro{ROS}[ROS]{Robot Operating System}
\newacro{TCP}[TCP]{Tool Center Point}
\newacro{DnD}[D\&D]{Design and Development}
\newacro{XR}[XR]{Extended Reality}
\newacro{CSV}[CSV]{Comma-separated values}
\newacro{PCA}[PCA]{Principal Component Analysis}
\newacro{RGBD}[RGB-D]{Red-Green-Blue-Depth}

\AtBeginDocument{%
  \providecommand\BibTeX{{%
    \normalfont B\kern-0.5em{\scshape i\kern-0.25em b}\kern-0.8em\TeX}}}

\setcopyright{rightsretained}
\acmJournal{PACMHCI}
\acmYear{2024} 
\acmVolume{8} 
\acmNumber{EICS} 
\acmArticle{244} 
\acmMonth{6}
\acmDOI{10.1145/3660243}

\makeatletter
\gdef\@copyrightpermission{
  \begin{minipage}{0.3\columnwidth}
   \href{https://creativecommons.org/licenses/by/4.0/}{\includegraphics[width=0.90\textwidth]{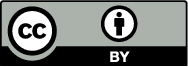}}
  \end{minipage}\hfill
  \begin{minipage}{0.7\columnwidth}
   \href{https://creativecommons.org/licenses/by/4.0/}{This work is licensed under a Creative Commons Attribution International 4.0 License.}
  \end{minipage}
  \vspace{5pt}
}
\makeatother

\newcommand\change[1]{{#1}}

\begin{document}

\title[AdaptiX -- A Transitional XR Framework for Assistive Robotics]{AdaptiX -- A Transitional XR Framework for Development and Evaluation of Shared Control Applications in Assistive Robotics}

\author{Max Pascher}
\orcid{0000-0002-6847-0696}
\email{max.pascher@udo.edu}
\affiliation{
    \institution{TU Dortmund University}
    \city{Dortmund}
    \country{Germany}
}
\affiliation{
    \institution{University of Duisburg-Essen}
    \city{Essen}
    \country{Germany}
}

\author{Felix Ferdinand Goldau}
\orcid{0000-0003-4552-6842}
\email{felix.goldau@dfki.de}
\affiliation{
    \institution{German Research Center for Artificial Intelligence (DFKI)}
    \city{Bremen}
    \country{Germany}
}

\author{Kirill Kronhardt}
\orcid{0000-0002-0460-3787}
\email{kirill.kronhardt@udo.edu}
\affiliation{
    \institution{TU Dortmund University}
    \city{Dortmund}
    \country{Germany}
}


\author{Udo Frese}
\orcid{0000-0001-8325-6324}
\email{udo.frese@dfki.de}
\affiliation{
    \institution{German Research Center for Artificial Intelligence (DFKI)}
    \city{Bremen}
   \country{Germany}
}

\author{Jens Gerken}
\orcid{0000-0002-0634-3931}
\email{jens.gerken@udo.edu}
\affiliation{
    \institution{TU Dortmund University}
    \city{Dortmund}
    \country{Germany}
}

\authorsaddresses{%
Authors’ addresses: 
\href{https://orcid.org/0000-0002-6847-0696}{Max Pascher}, TU Dortmund University, Emil-Figge-Str. 50. 44227 Dortmund, Germany, \href{mailto:max.pascher@udo.edu}{max.pascher@udo.edu}; 
\href{https://orcid.org/0000-0003-4552-6842}{Felix Ferdinand Goldau}, Enrique-Schmidt-Str. 5, 28359 Bremen, Germany, \href{mailto:felix.goldau@dfki.de}{felix.goldau@dfki.de}; 
\href{https://orcid.org/0000-0002-0460-3787}{Kirill Kronhardt}, TU Dortmund University, Emil-Figge-Str. 50. 44227 Dortmund, Germany, \href{mailto:kirill.kronhardt@udo.edu}{kirill.kronhardt@udo.edu}; 
\href{https://orcid.org/0000-0001-8325-6324}{Udo Frese}, Enrique-Schmidt-Str. 5, 28359 Bremen, Germany, \href{mailto:udo.frese@dfki.de}{udo.frese@dfki.de}; 
\href{https://orcid.org/0000-0002-0634-3931}{Jens Gerken}, TU Dortmund University, Emil-Figge-Str. 50. 44227 Dortmund, Germany, \href{mailto:jens.gerken@udo.edu}{jens.gerken@udo.edu}.
}

\renewcommand{\shortauthors}{M. Pascher et al.}

\begin{abstract}

With the ongoing efforts to empower people with mobility impairments and the increase in technological acceptance by the general public, assistive technologies, such as collaborative robotic arms, are gaining popularity.  
Yet, their widespread success is limited by usability issues, specifically the disparity between user input and software control along the autonomy continuum. 
To address this, shared control concepts provide opportunities to combine the targeted increase of user autonomy with a certain level of computer assistance.
This paper presents the free and open-source \emph{AdaptiX} \acs{XR} framework for developing and evaluating shared control applications in a high-resolution simulation environment. The initial framework consists of a simulated robotic arm with an example scenario in \acf{VR}, multiple standard control interfaces, and a specialized recording/replay system. \emph{AdaptiX} can easily be extended for specific research needs, allowing \ac{HRI} researchers to rapidly design and test novel interaction methods, intervention strategies, and multi-modal feedback techniques, without requiring an actual physical robotic arm during the early phases of ideation, prototyping, and evaluation. Also, a \ac{ROS} integration enables the controlling of a real robotic arm in a \emph{PhysicalTwin} approach without any simulation-reality gap. 
Here, we review the capabilities and limitations of \emph{AdaptiX} in detail and present three bodies of research based on the framework.  
\emph{AdaptiX} can be accessed at \change{\url{https://adaptix.robot-research.de}}.
\end{abstract}




\begin{teaserfigure}
\centering
    \subfloat[]{\includegraphics[width=0.325\linewidth]{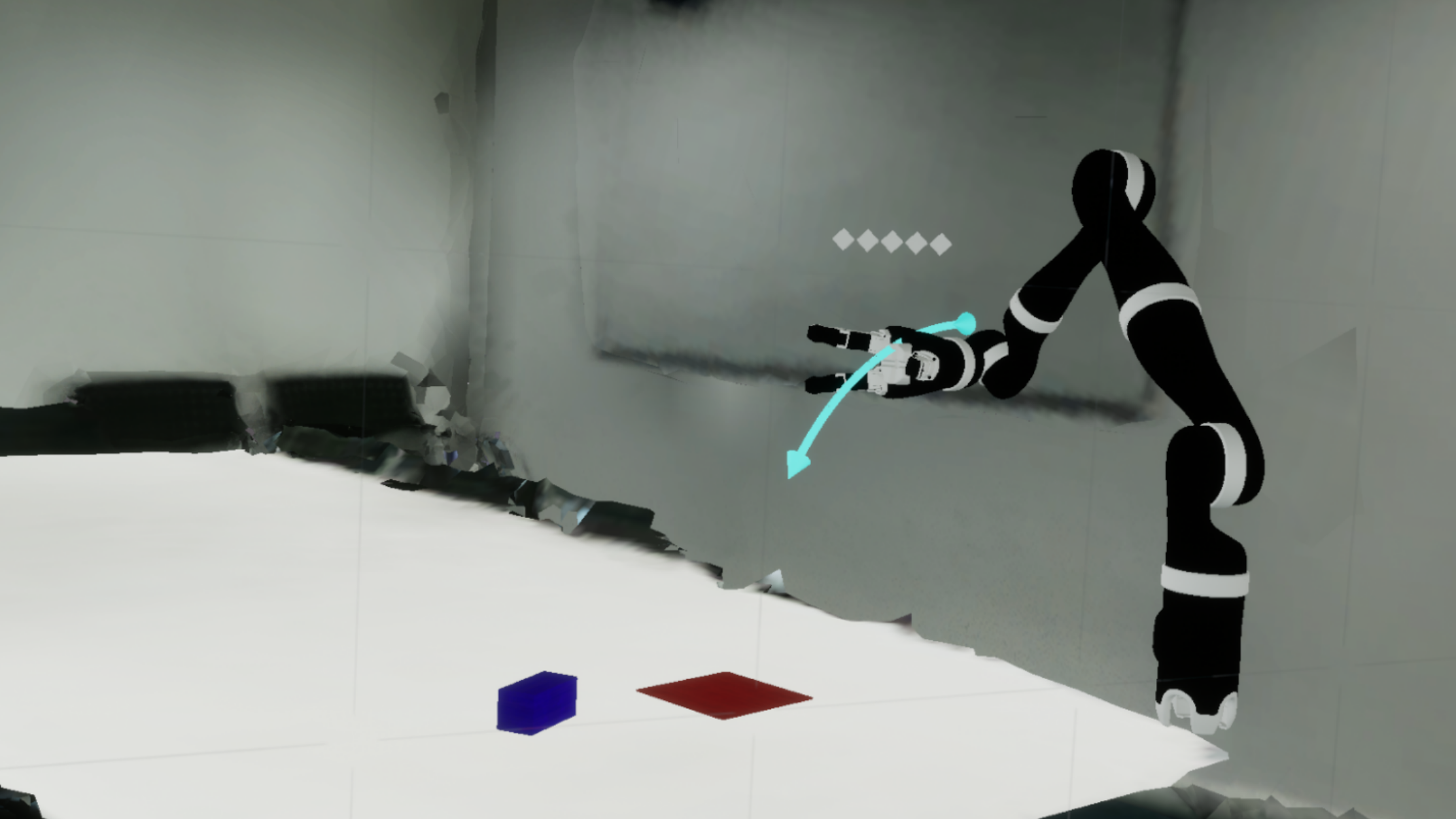}\label{fig:mr-setting_a}}
    \hfill
    \subfloat[]{\includegraphics[width=0.325\linewidth]{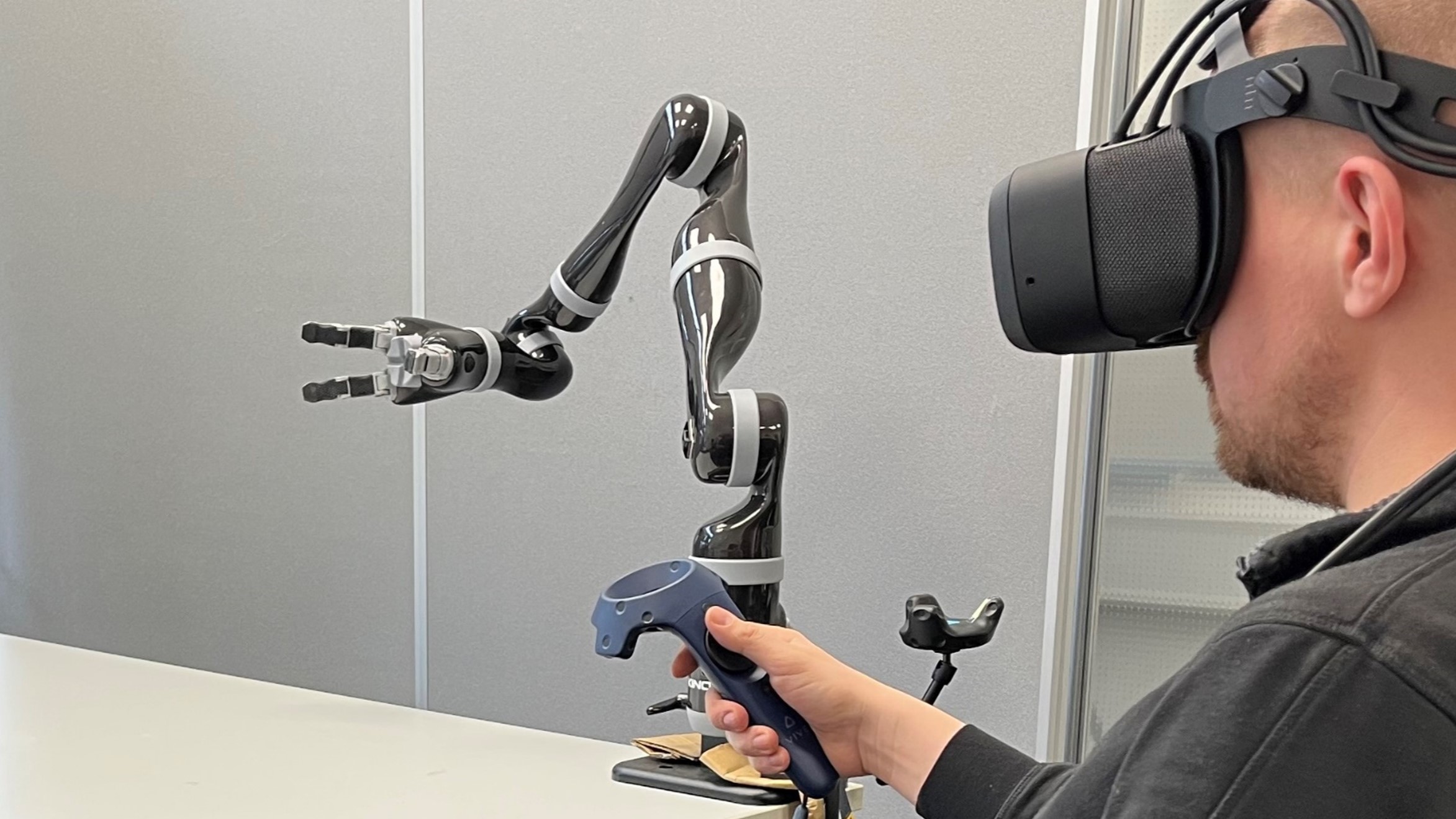}\label{fig:mr-setting_b}}
    \hfill
    \subfloat[]{\includegraphics[width=0.325\linewidth]{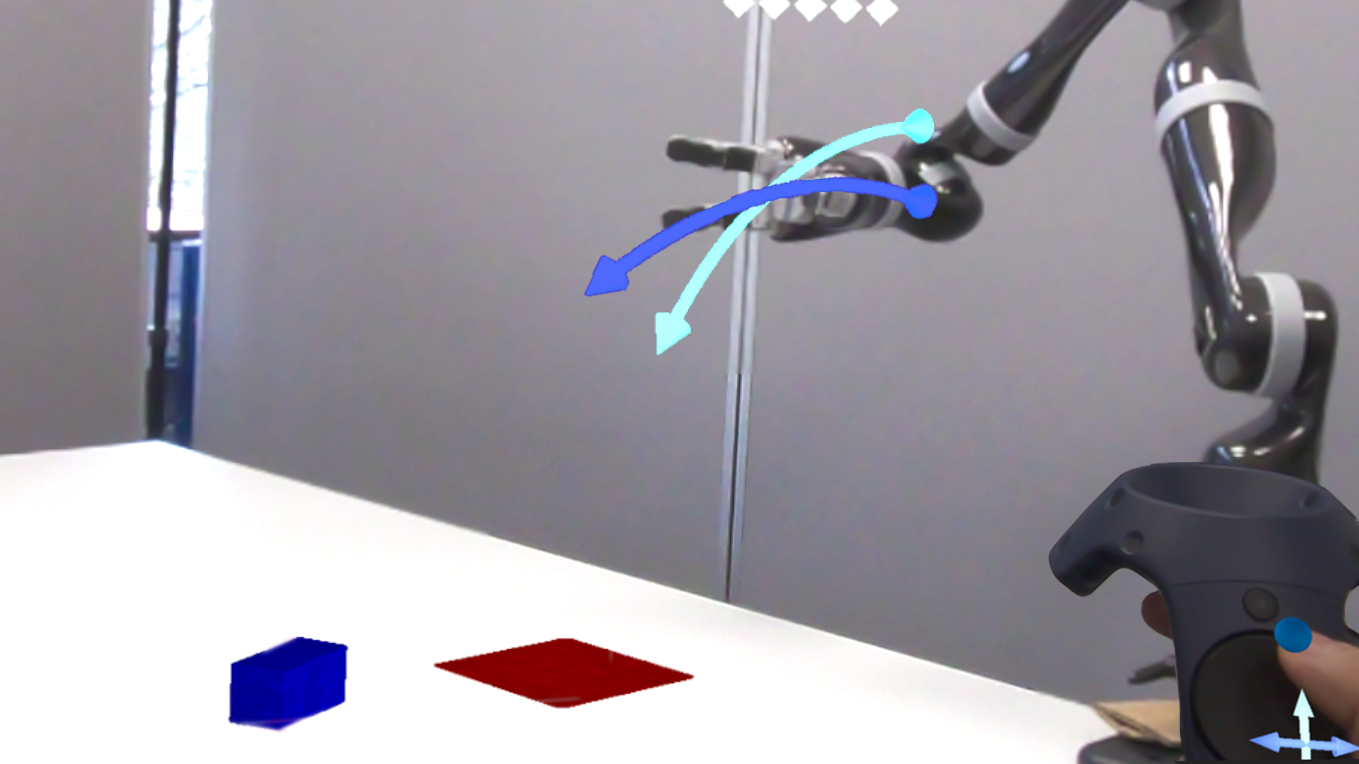}\label{fig:mr-setting_c}}
\caption{Setup with (\textbf{a}) a user's view in the \ac{VR} simulation environment, (\textbf{b}) setup of interaction with a physical robot, and (\textbf{c}) a combined view of physical robot and visual cues in \ac{MR}.}
\label{fig:teaser}
\end{teaserfigure}

\maketitle

\section{Introduction}
\label{sec:introduction}

\change{
Robotic arms as assistive technologies are a powerful tool to increase self-sufficiency in people with limited mobility~\cite{Kyrarini.2021survey,Pascher.2021recommendations}, as they facilitate the performance of \ac{ADLs} -- usually involving grasping and manipulating objects in their environment -- without human assistance~\cite{Petrich.2022ADL}. 
However, a frequent point of contention is the assistive robot's autonomy level. The reduction of user interaction to just oversight with purely autonomous systems elicits stress~\cite{Pollak.2020} and feelings of distrust in their users~\cite{zlotowski2017can}. 
On the other side of the autonomy spectrum, manual controls can be challenging - or even impossible - to operate, depending on the significance and type of impairment. Shared control -- a combination of manual user control through standard input devices plus algorithmic support through computer software adjusting the resulting motion -- may have the potential to mitigate both concerns~\cite{Abbink.2018}. Here, both the user and the robot share a task on the operational level, enabling people with motor impairments to get involved in their assistance. As a result, such approaches can increase the feeling of independence while improving ease of use compared to manual controls~\cite{Flemisch2019}.
}

\change{A characteristic real-world scenario, motivated by our research, has an assistive robotic arm (e.g., a Kinova Jaco 2) attached to a wheelchair to support the user in \ac{ADLs}. Here, the user is challenged with operating six or more \acp{DoF}, which requires complex input devices or time-consuming and confusing mode switches. This potentially results in increased task completion time and user frustration~\cite{Herlant.2016modeswitch}. Addressing this, shared control systems can facilitate more straightforward and accessible robot operation. However, they may require well-designed communication of robot (motion) intent, so that the user retains awareness and understands the level of support they get from the system~\cite{Pascher.2023robotMotionIntent}. Also, different users might need distinct input devices or require multi-modal input to account for varying abilities. 

Based on our experiences, we identified several challenges that currently influence and potentially impede the effective development of shared control approaches:
\begin{itemize}
    \item Shared control systems for assistive technologies still pose open questions requiring considerable experimentation, tweaking and balancing between user and robot interaction~\cite{Lebrasseur.2019}.
    \item While much research explored robot motion intent, there is little insight into what works best in which situation and for which type of user. In assistive robotics, the visualization and feedback modality must be carefully adapted to the user's needs and abilities as there is no \enquote{one size fits all} solution~\cite{Holloway.2019onesizefitsone}.
    \item Similarly, suitable input devices may vary between users. Depending on individual preferences and capabilities, multi-modal input or the choice between different input modalities may be required~\cite{Arevalo-Arboleda2021b}.
    \item Bringing robots and humans physically together during research studies is difficult due to the laborious and costly transportation, safety concerns with robots and general availability of the user group~\cite{Bustamante.2021challenge}. 
\end{itemize}
}

\textbf{Contribution.} To \change{allow researchers, designers and developers to} address th\change{e}s\change{e challenges holistically and flexibly, we present} \emph{AdaptiX} -- a free, open-source \acs{XR} framework \change{\footnote{\emph{AdaptiX} framework. \url{https://adaptix.robot-research.de}, last retrieved \today.}}. \change{Aimed at \ac{DnD}, \emph{AdaptiX} combines a physical robot implementation with a 3D simulation environment. The simulation approach (analogous to simulations in industrial settings~\cite{Matsas.2015vrIndustry,Muller.2017vrIndustry,Tidoni.2017vrIndustry}) mitigates the assistive robotic arm's bulky, expensive, and complex nature. It also makes the integration of visualization feedback or different input modalities easier to explore and test, while a \acf{ROS} interface allows the direct transfer to the real robot. Testing new interaction and control options becomes much less time-consuming while simultaneously excluding potentially dangerous close-contact situations with users before glitches are managed~\cite{Pascher.2021recommendations}. In total, the framework facilitates the development and evaluation of assistive robot control applications \emph{in-silico} and} creates a practical and effective step between ideation, development, and evaluation, allowing \ac{HRI} researchers more flexibility and facilitating efficient resource usage.

To summarize, the \emph{AdaptiX} framework contributes the following:

\begin{itemize}
    \item \emph{AdaptiX} allows researchers to rapidly design and test novel visualization and interaction methods.

    \item The framework integrates an initial concept and implementation of a shared control approach.

    \item The integrated \ac{ROS} interface facilitates connection to a non-simulated -- physical -- robotic arm to perform bidirectional interactions and data.

    \item The framework's concept enables a code-less trajectory programming by hand-guiding the simulated or physical assistive robotic arm to the specific location and saving the position and orientation of the \ac{TCP}.

    \item Recording \ac{TCP} data enables replaying user-controlled robot movements and results in a fully customizable system. Options include changing specific details during replaying, such as repositioning cameras or re-rendering background scenes. 
    
    \item Finally, the entire continuum of \acf{MR} can be exploited in the \emph{AdaptiX} environment. This allows applications in \change{\acf{VR}, pure screen space}, \ac{AR}, simultaneous simulation and reality, and pure reality (cf. the \emph{virtuality continuum} of \citet{milgram1994taxonomy}).

\end{itemize}

\section{Related Work}
\label{sec:related-work}

While robotic arms are a particularly useful and versatile subset of assistive technologies, their widespread success is limited by a number of design challenges concerning the interaction with their human user. In recent years, a growing body of research addressed these concerns and associated optimization options to increase their usability, e.g.,~\cite{Lebrasseur.2019,Cio.2019,Haseeb.2018}. During the \emph{AdaptiX} development process, we aimed to include functionality to address the challenges of shared control optimization~\cite{Goldau.2021petra}, intent communication~\cite{Pascher.2023robotMotionIntent}, and attention guidance~\cite{PETERMEIJER2017204}.  

\subsection{Shared Control for Assistive Robots}
\label{sec:current-shared-controls}

Current shared control systems operate along an autonomy continuum, respectively balancing user input and system adjustments. At one extreme, the systems tend to be heavily manual, with only minor adjustments to the user's input~\cite{Sijs.2007}. On the other end are systems where users primarily provide high-level commands for the robot to execute~\cite{tsui2011want}. A number of different approaches -- including time-optimal~\cite{Herlant.2016modeswitch} and blended mode switching~\cite{ezeh2017}, shared-control-templates~\cite{quere2020} and body-machine-interfaces~\cite{jain2015} -- are currently employed in various settings. 

A fundamentally different approach is the shared control system proposed by \citet{Goldau.2021petra}. Their concept combines a robotic arm's cardinal \acp{DoF} according to the current situation and maps them to a low-\ac{DoF} input device. The mapping is accomplished by attaching a camera to the robotic arm's gripper and training a \ac{CNN} by having people without motor impairments perform \ac{ADLs}~\cite{Goldau.2021petra} -- similar to the learning-by-demonstration approach for autonomous robots by \citet{Canal.2016}. The \ac{CNN} returns a set of newly mapped \acp{DoF}, ranked by their assumed likeliness based on the \ac{CNN} for the given situation, allowing users to access a variety of movements for each situation. 
In addition, the \ac{CNN}-based approach allows the system to be easily extendable as the same system can be trained to discriminate between many different situations -- making it a viable concept for day-to-day use. \citet{Goldau.2021petra} conducted a proof-of-concept study comparing the control of a simulated 2D robot with manual or \ac{CNN}-based controls. Task execution was faster with their proposed concept; however, users experienced it as more complex than manual controls~\cite{Goldau.2021petra}.

Our framework \emph{AdaptiX} is influenced by \citeauthor{Goldau.2021petra}'s approach, but extends it from 2D to 3D space. This increases the number of possible \acp{DoF}, which allows for an accurate representation of \ac{ADLs} in the framework. By adding functionality, visualizations, and a \ac{ROS} integration, \emph{AdaptiX} can be used to develop and evaluate novel interaction control methods based on this approach for shared control, which we refer to as \emph{\ac{ADMC}}.

\subsection{Robot Motion Intent}
\label{sec:background-robotIntents}

Regardless of the specific interaction details, it is necessary to effectively communicate the intended assistance provided by the (semi-)autonomous system~\cite{brooks2020}. Clear communication between robots and humans enhances the shared control system's predictability, avoids accidents, and increases user acceptance.

A crucial element of the \ac{DnD} process of robotic devices is, therefore, the testing of intent communication methods. 
\change{\emph{Choreobot} -- an interactive, online, and visual dashboard -- proposed by \citet{Deurzen.2022} supports researchers and developers to identify where and when adding intelligibility to the interface design of a robotic system improves the predictability, trust, safety, usability, and acceptance.
Moreover, }\citet{Pascher.2023robotMotionIntent} provide an extensive overview of the various types of visualization and modalities frequently used in communicating robot motion intent. These range from auditory~\cite{Chen.2011} and haptic~\cite{che2020} modalities to anthropomorphizing the robot and using its gaze~\cite{may2015} or gestures~\cite{gielniak2011generating}. Their findings are substantiated by \citet{Holthaus.2023.typology}, who used an ethnographic approach to derive a comprehensive communication typology. 

While all these intent communication modalities are viable, visual representations of future movements are often quoted as less workload-intense for the end-user~\cite{cleaver_2021}. \ac{AR} is, therefore, unsurprisingly a frequently used tool to convey detailed motion intent~\cite{walker_2018_armotionintent, ruffaldi2016third, Hetherington_2021, chadalavada2020bi, zein2021deep}, allowing interactions to become more intuitive and natural to humans~\cite{makhataeva2020}. \citeauthor{Suzuki.2022.AR-HRC-Survey} emphasize the benefits of \ac{AR}-based visualizations for communicating movement trajectories or the internal state of the robot~\cite{Suzuki.2022.AR-HRC-Survey}. 

The visual feedback employed by \emph{AdaptiX} mimics \ac{AR} in a \ac{VR} environment with directional cues registered in 3D space. This approach allows the user to understand different movement directions for the actual control and the suggested \ac{DoF} combinations. To streamline understanding the control methods, one of our primary approaches is the usage of arrows -- a straightforward and common visualization technique to communicate motion intent~\cite{walker_2018_armotionintent, shrestha_2016_motionintent, shindev_2012_intentexpression}.

\subsection{Feedback Modalities for User Attention Guidance}
\label{sec:attention-guidance}
When creating systems using shared control, it is crucial to guide the user's focus to the assistance the robot is offering~\cite{petermeijer2017}. This guidance is particularly important if either party is moving the robot in a way that could lead to collisions or worsen the situation. To enhance the predictability of shared control systems, various feedback modalities have been proposed to guide user attention as a secondary feedback mechanism to \ac{AR}. The goal is to provide a feedback solution that results in short reaction times, enabling users to quickly direct their focus to the information provided by the robot.

In the related discipline of autonomous driving systems, if the vehicle encounters a situation it was not programmed or trained to handle, it will issue a \ac{TOR}. This \ac{TOR} prompts the driver to take manual control of the vehicle to prevent a collision or to drive in areas the vehicle cannot handle autonomously.

Auditory, visual, and tactile/haptic modalities are commonly used for \acp{TOR}~\cite{yun2020multimodal} -- either as a single sensory input~\cite{petermeijer2017} or a combination of multiple variants~\cite{PETERMEIJER2017204}. Simulation studies, along with research on reaction times to different sensory stimuli, indicate that multi-modal feedback results in the lowest possible reaction times in shared control systems~\cite{burke2006comparing, diederich2004bimodal, kosinski2008literature}. 

\begin{figure}[htbp]
    \centering
    \includegraphics[width=\linewidth]{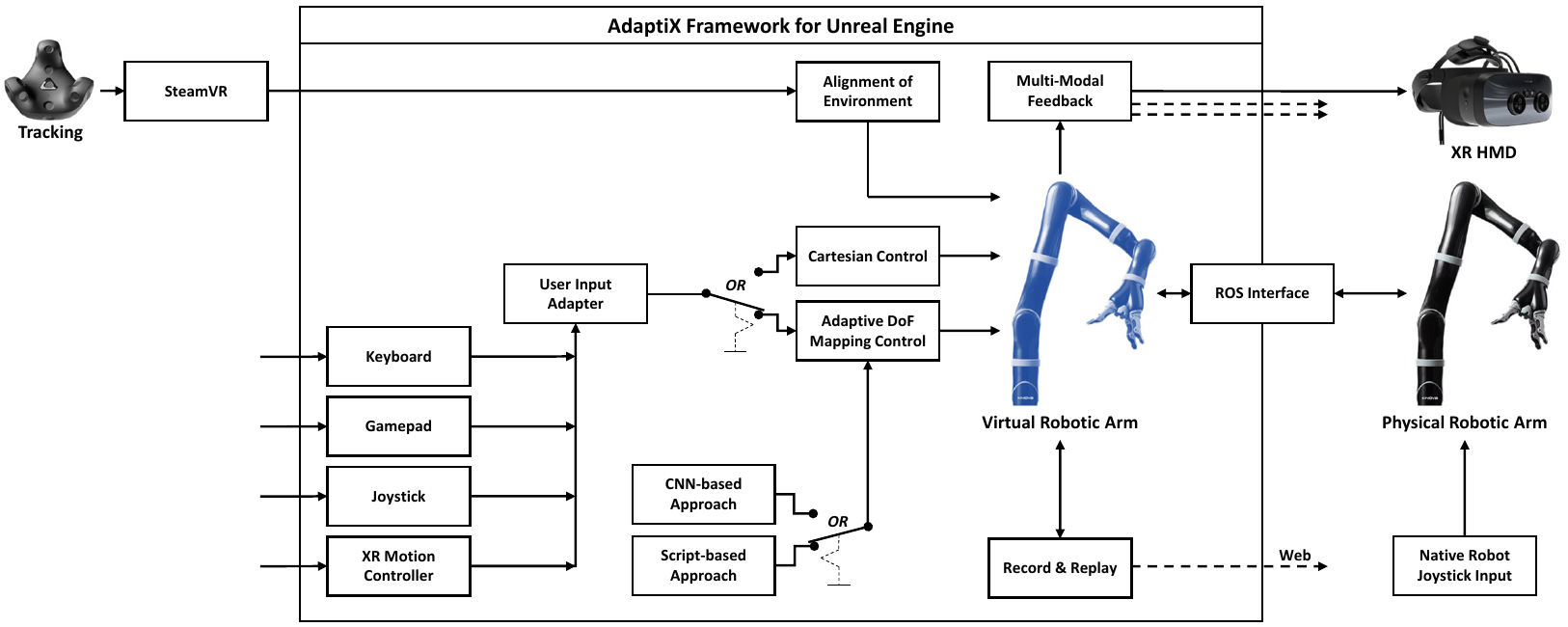}
    \caption{Overview of \emph{AdaptiX}' architecture, illustrating each component, their directional communication, and the crossover from and to the framework. The user input is either used for \emph{Cartesian Control} or \acf{ADMC}. For \ac{ADMC}, either a \ac{CNN}-based or script-based rule engine can be selected.}
    \label{fig:framework-architecture}
\end{figure}

Implementing these feedback methods into existing assistive robot systems would be straightforward as the necessary output devices -- like screens, speakers, or vibration motors -- are commonly already present. To allow researchers to evaluate the benefits of the different modalities, \emph{AdaptiX} includes three modes for attention guiding: visual, auditory, and tactile/haptic. Developers can either choose one modality or follow a multi-modal approach.
\change{ 

\section{Framework Concept}
\label{sec:concept}

The \emph{AdaptiX} \acs{XR} framework facilitates the development and evaluation of \ac{HRI} shared control applications in an easy-to-use, high-resolution transitional \ac{MR} environment. Equipped with a \ac{VR} simulation environment containing a virtual \emph{Kinova Jaco 2} and ample customization options, researchers can streamline their \ac{DnD} process while simultaneously reducing overhead and boosting efficiency. 
\autoref{fig:framework-architecture} provides an overview of the framework's architecture.

} 

\change{ 

In addition to an Cartesian robot control, we propose \ac{ADMC} as an initial shared control approach, using suggestions by a rule engine (e.g., a \ac{CNN} or script-based approach) to be controlled by the user. 
\ac{ADMC} is implemented directly into the \emph{Unreal Engine} to enable researchers and developers to fully customize the control methods, systems behavior, and feedback techniques by coding in \emph{C++} or \emph{Blueprints}. 

\emph{AdaptiX} supports several pre-implemented input devices and provides an adapter class for an easy development and implementation of further input devices. This supports researchers and developers to easy implement their ideas and concepts.
The integrated \ac{ROS} interface facilitates connection to a non-simulated -- physical -- robotic arm to perform bidirectional interactions and data exchange in a \emph{DigitalTwin} and \emph{PhysicalTwin} approach. 

\emph{AdaptiX} enables effortless trajectory programming by manually guiding the \ac{TCP} of a simulated or physical robotic arm to a desired location and recording its position and orientation. Recorded data of user-controlled robot movements can be replayed. Offering the adjustment of specific details, such as camera positions and background scenes, results in a highly customizable system.

The aim is to provide a modular and extensible framework so that research teams do not need to start from scratch when implementing their shared control applications.

} 

\subsection{Adaptive DoF Mapping Control (ADMC)} 
\label{sec:concept-adaptiveControl}

For the adaptive \ac{DoF} mapping -- referred to as \emph{\ac{ADMC}} -- of the robotic arm, the goal is to present a set of \ac{DoF} mappings ordered based on their effectiveness in accomplishing the pick-and-place task used in the experiment. The concept of \enquote{usefulness} assumes that maximizing the cardinal \acp{DoF} of the robot assigned to an input-\ac{DoF} while progressing towards the next goal is the most advantageous option.

This \ac{DoF} mapping, referred to as the \emph{optimal suggestion}, is assumed to be the best choice due to a significant reduction in the need for mode switches when multiple \acp{DoF} are combined into a single movement. The more \acp{DoF} are combined (assuming it is sensible for the given situation), the fewer mode switches are required. As a result, the \ac{DoF} mappings are ordered based on the number of \acp{DoF} they combine.

In addition to the optimal suggestion, the second suggestion \change{is a selection of an orthogonal variation of the first suggestion, which has the highest probability and most variation in spatial direction and keeps the number of combined \acp{DoF} unchanged}. This secondary suggestion is likely useful to users as they can utilize it to adjust their position while maintaining a sensible orientation toward the next goal. \change{The following \ac{DoF} mappings were used} (see \autoref{fig:adaptive-suggestions}):

\begin{figure}[htbp]
\centering
\captionsetup{justification=centering}
    \subfloat[]{\includegraphics[width=0.16\linewidth]{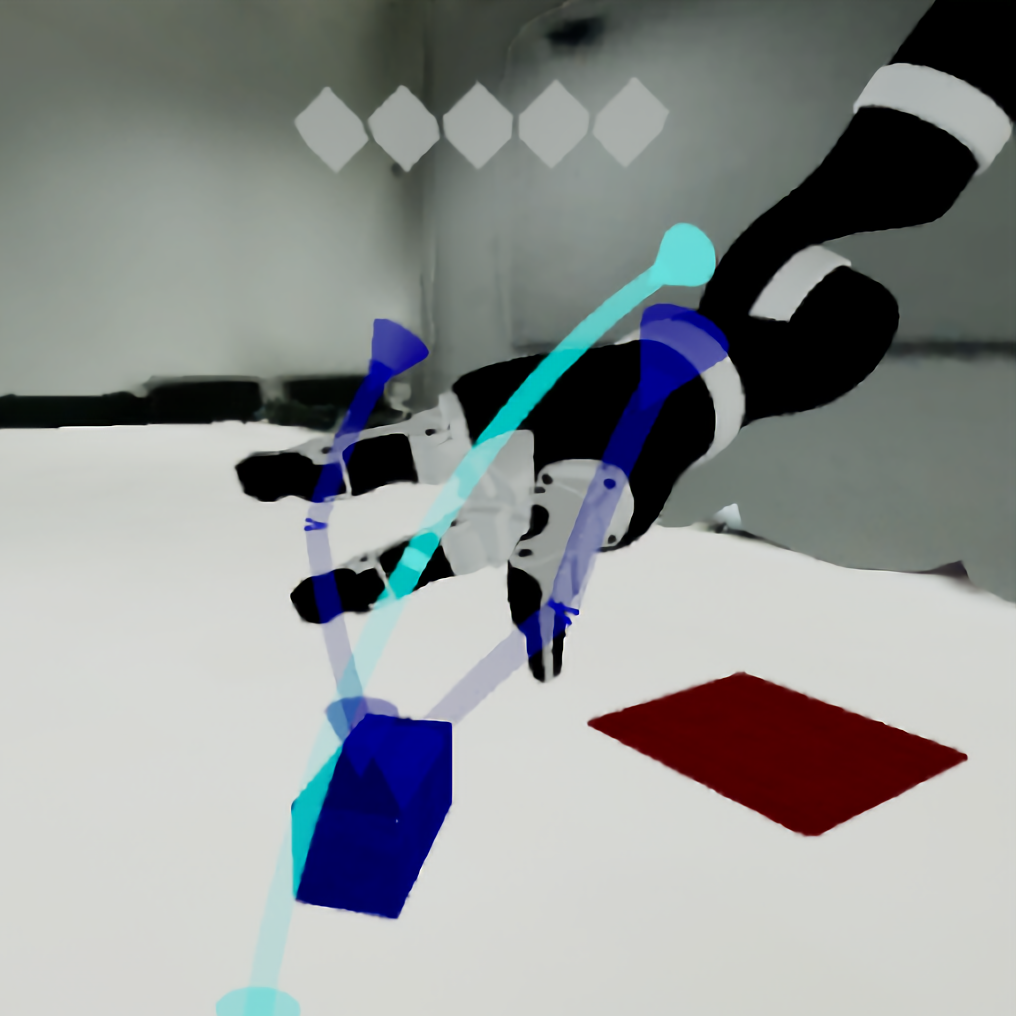}\label{fig:adaptive-suggestions:a}}
    \hfill
    \subfloat[]{\includegraphics[width=0.16\linewidth]{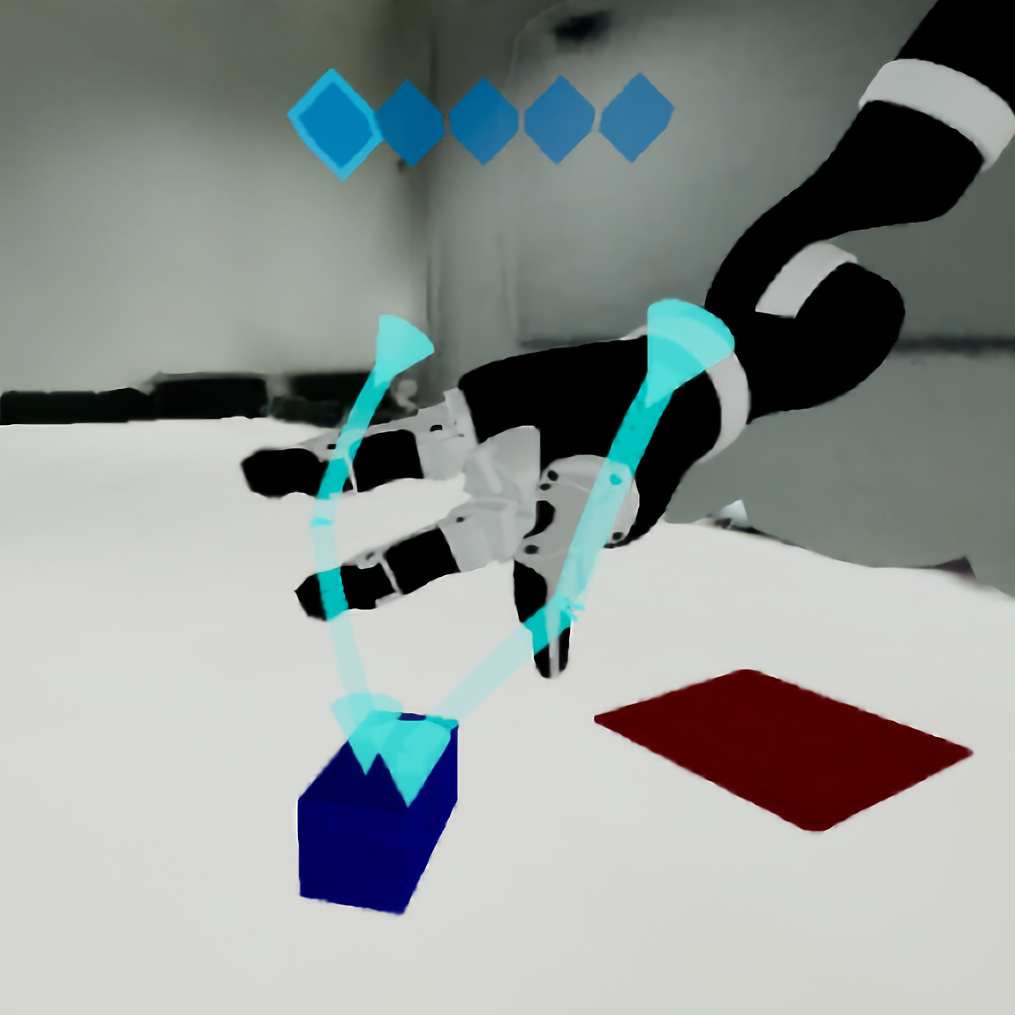}\label{fig:adaptive-suggestions:b}}
    \hfill
    \subfloat[]{\includegraphics[width=0.16\linewidth]{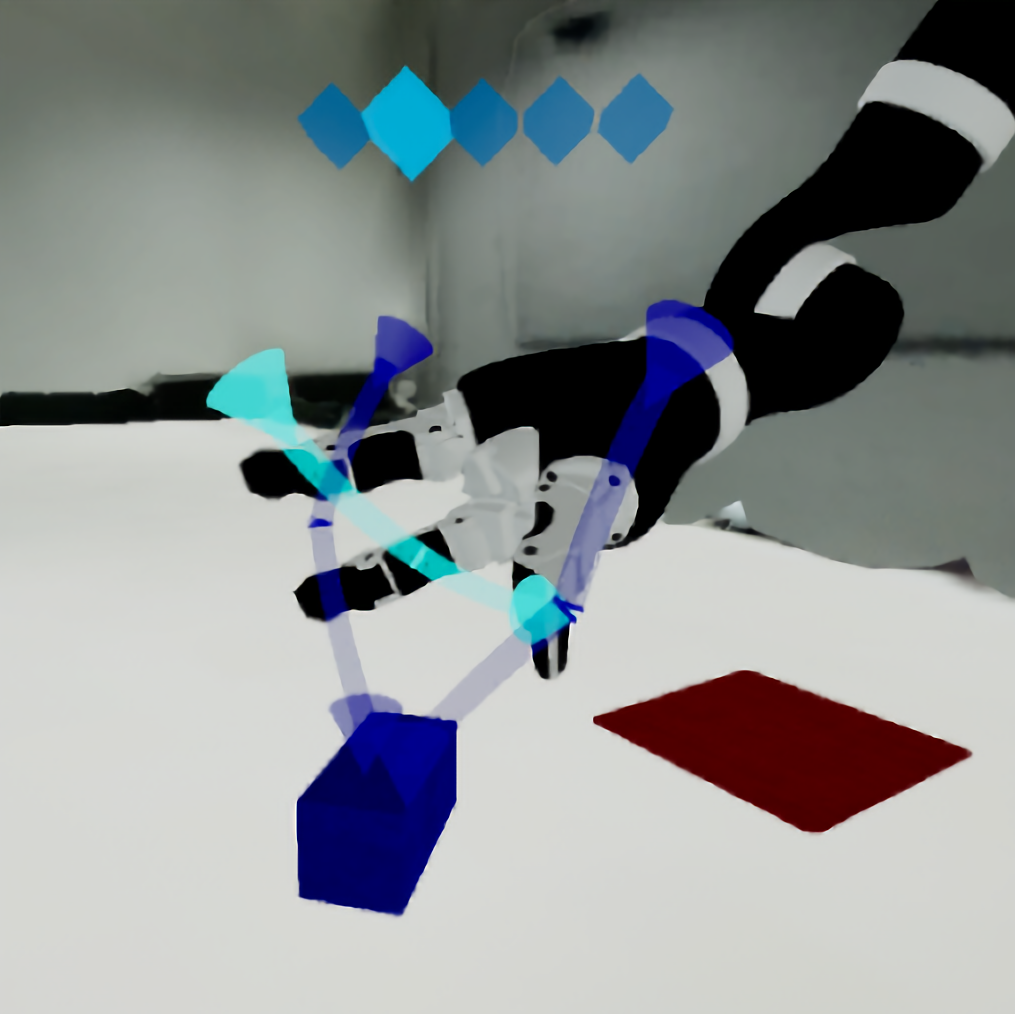}\label{fig:adaptive-suggestions:c}}
    \hfill
    \subfloat[]{\includegraphics[width=0.16\linewidth]{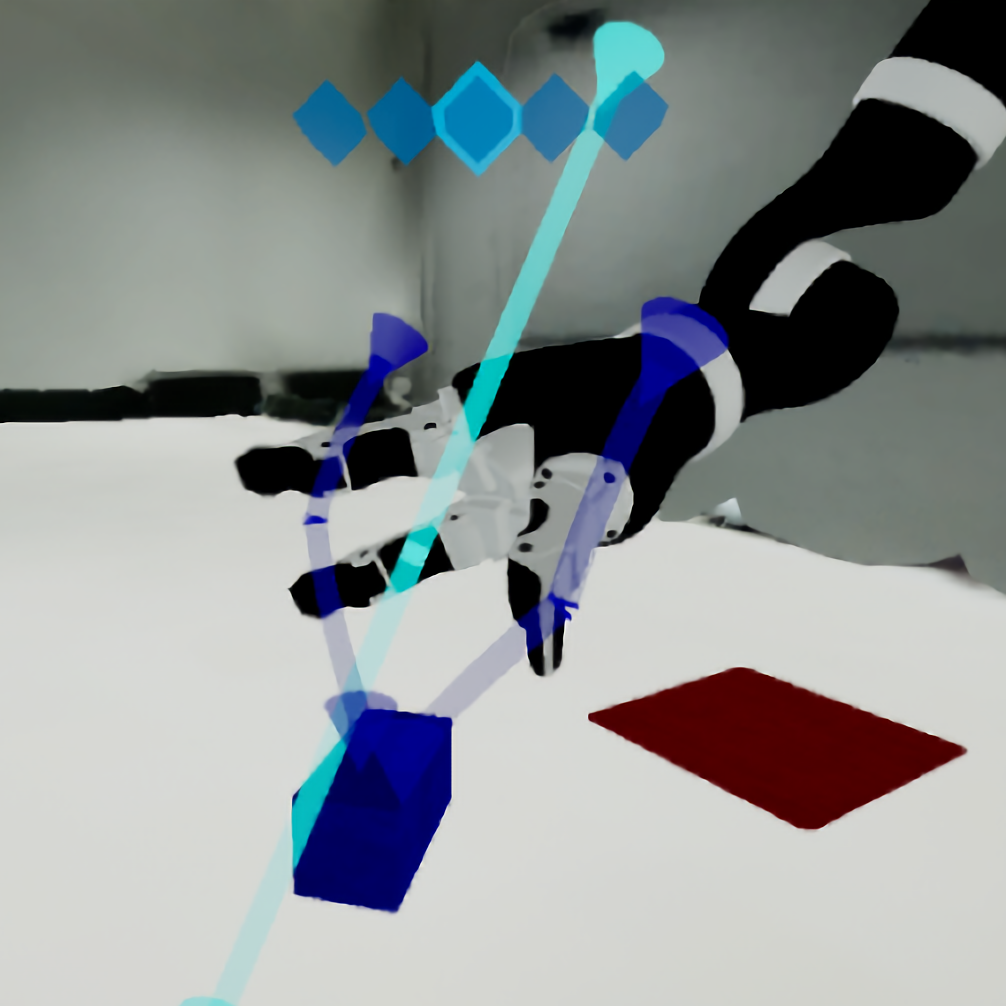}\label{fig:adaptive-suggestions:d}}
    \hfill
    \subfloat[]{\includegraphics[width=0.16\linewidth]{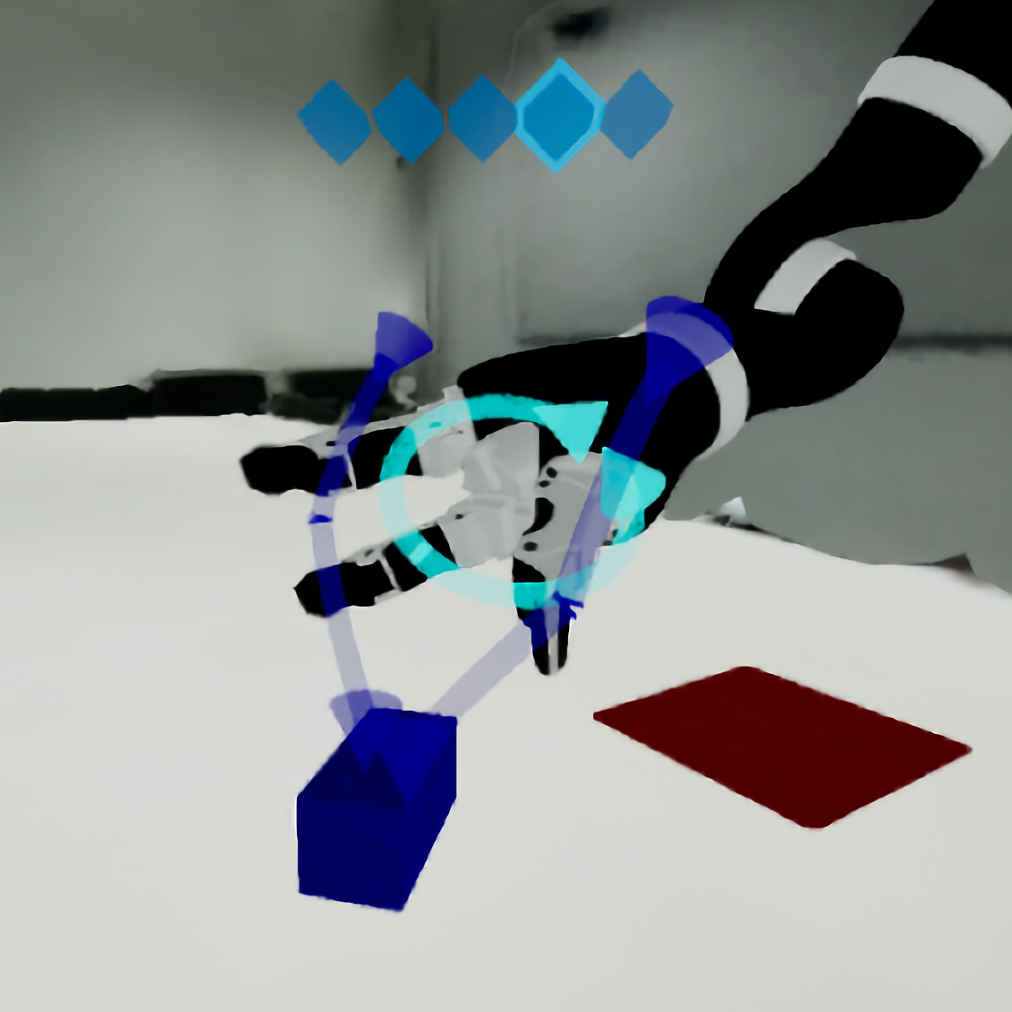}\label{fig:adaptive-suggestions:e}}
    \hfill
    \subfloat[]{\includegraphics[width=0.16\linewidth]{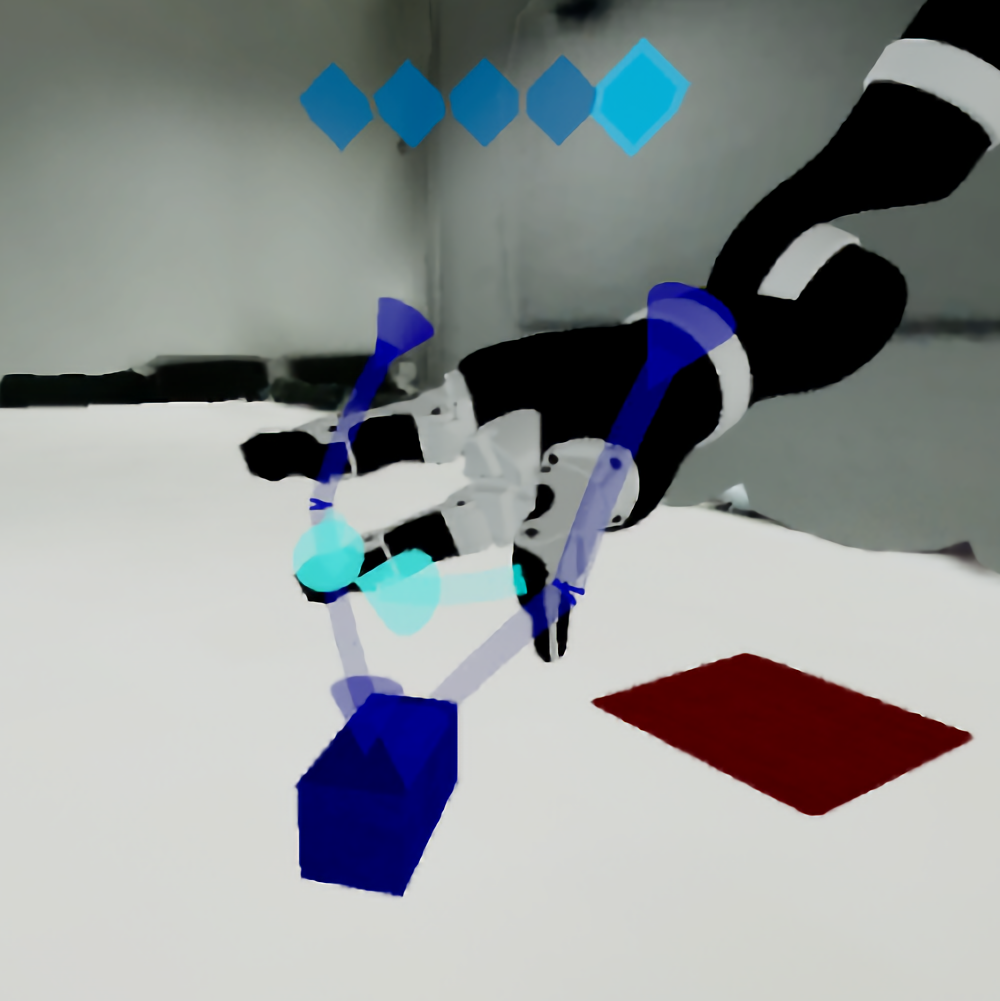}\label{fig:adaptive-suggestions:f}}
\captionsetup{justification=justified}
  \caption{Suggestions as Visualized in the \ac{ADMC}, (\textbf{a}) Continue previous movement, (\textbf{b}) Optimal Suggestion, (\textbf{c}) Adjustment Suggestion, (\textbf{d}) Translation Suggestion, (\textbf{e}) Rotation Suggestion, (\textbf{f}) Gripper Suggestion. Colors: Bright cyan arrow: Currently active \ac{DoF} mapping. Dark blue arrow: Next most likely \ac{DoF} mapping.}
  \label{fig:adaptive-suggestions}
\end{figure}

\begin{enumerate}
    \item \emph{Optimal Suggestion:} Combining translation, rotation, and finger movement [opening and closing] into one suggestion, causing the gripper to move towards the target, pick it up, or release it on the intended surface.
    \item \emph{Adjustment Suggestion:} An orthogonal suggestion based on (1) but excluding the finger movement. Allows the users to adjust the gripper's position while still being correctly orientated.
    \item \emph{Translation Suggestion:} A pure translation towards the next target, disregarding any rotation.
    \item \emph{Rotation Suggestion:} A pure rotation towards the next target disregarding any translation.
    \item \emph{Gripper Suggestion:} Opening or closing of the gripper's fingers.
\end{enumerate}

\subsubsection{\ac{CNN}-based Approach}
For the \ac{CNN} approach, a color-and-depth camera is attached to the gripper of an assistive robotic arm. The live video feed is transmitted to a \ac{CNN}, which is trained using data collected from non-impaired individuals performing \ac{ADLs} using the robotic arm along with a high-\ac{DoF} input device. \change{The \ac{CNN} does not need a model of the environment to provide these mappings.} \ac{PCA} is employed to transform the \ac{CNN}'s output into a matrix \emph{\^D}, where each column represents a combination of cardinal \acp{DoF} along which the robotic arm can move.

\begin{figure}[htbp]
\centering
\subfloat[]{\includegraphics[width=0.635\linewidth]{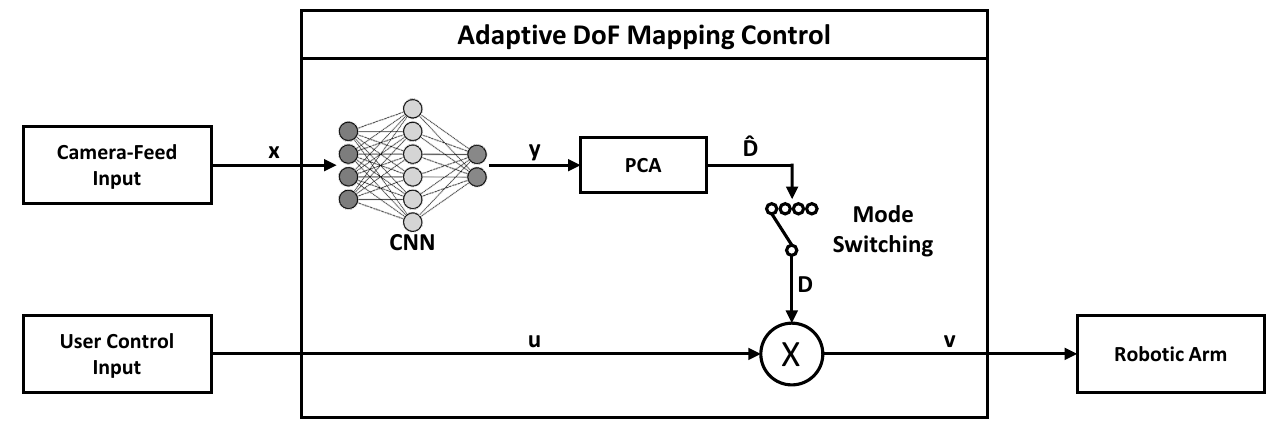}\label{fig:goldau-pipeline}}
  \hfill
  \subfloat[]{\includegraphics[width=0.365\linewidth]{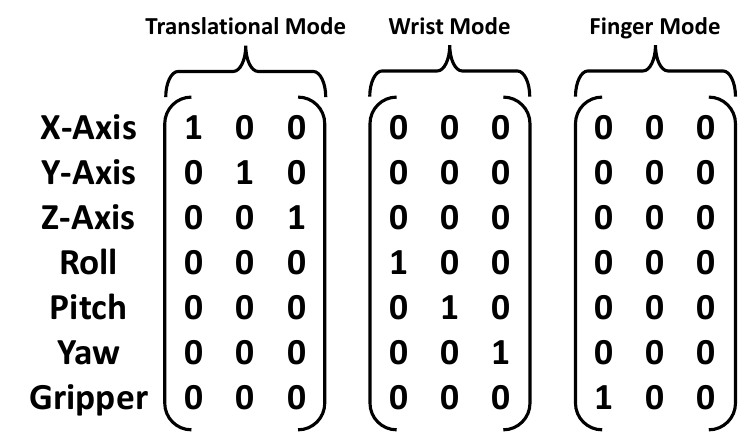}\label{fig:goldau-matrix}}
\centering
\caption{Concept of adaptive \ac{DoF} mapping control. \textbf{(a)} Control pipeline for proposed adaptive shared control and \textbf{(b)} matrix representation of \ac{DoF} mappings: Columns represent input-\acp{DoF}. Rows represent output-\acp{DoF}. Subsets represent modes. Two empty columns were added to represent zero movement mappings in \emph{Finger Mode}.}
\label{fig:admc-concept}
\end{figure}

Next, a subset of \emph{\^D} is selected, containing as many columns as the number of \acp{DoF} provided by the input device. This selected subset is referred to as \emph{D}, and it serves to map input-\acp{DoF} to output-\acp{DoF}. When an input-\ac{DoF} is engaged, the robot's movements are determined by the values in the corresponding vector of \emph{D}, which proportionally activate the robot's cardinal \acp{DoF}.
A mode switch is defined as the exchange of \emph{D} with a different subset of \emph{\^D}. This enables the system to switch between various mappings of input-\acp{DoF} to output-\acp{DoF}, adapting the robot's control according to the user's needs and preferences.
A visual representation of this control pipeline is depicted in \autoref{fig:goldau-pipeline}.

\emph{\^D} is a square matrix with dimensions based on the number of cardinal \acp{DoF} available on the robot to be controlled. In the case of the \emph{Kinova Jaco 2}~\change{\cite{kinova}}, this results in a $7 \times 7$ matrix. This matrix represents a mapping of input-\acp{DoF} to output-\acp{DoF} when the number of \acp{DoF} in both cases is equal. The values in each column, ranging from -1 to 1, indicate the proportion with which the specific cardinal \ac{DoF} is utilized when engaging the corresponding input-\ac{DoF}.

By defining \emph{\^D} as an identity matrix, each input-\ac{DoF} is mapped to a single output-\ac{DoF}. Selecting an equal number of columns from \emph{\^D} to form matrix \emph{D} allows for manual control with mode switching along cardinal \acp{DoF}. Moreover, this representation enables the combination of multiple cardinal movements into arbitrary output \ac{DoF} mappings. For example, a (transposed) column of \emph{(0.5, 0.5, 0, 0, 0, 0, 0)} would result in diagonal movement along the X- and Y-Axes of the robot. Such combinations enable the offering of complex movements with different proportions depending on the situation, enhancing the control options available to users.
The identity matrix for a \emph{Kinova Jaco 2} with a 3-\acp{DoF} joystick is illustrated in \autoref{fig:goldau-matrix}. 

\subsubsection{Script-based Approach}
As an alternative rule engine for our \ac{ADMC} concept, we implemented a task-specific script. This approach eliminates potential biases that a more generic, but currently limited method like a \ac{CNN}-based control might introduce. It is essential to note that our task-specific script is effective only in a controlled experimental environment.

The task-specific script assesses the end effector's current position, rotation, and finger position relative to a target, allowing it to adaptively calculate the matrix \emph{\^D}. This script recommends optimal movements to pick up an object and place it onto a target drop area, maximizing the combination of as many \acp{DoF} as possible. Additionally, it provides other \ac{DoF} combinations that may be less beneficial to mimic the idea that each subsequent column in \emph{\^D} has a decreasing likelihood of being useful. These additional \ac{DoF} mappings are ordered by the number of combined \acp{DoF} in a decreasing manner.

To validate the effectiveness of this approach, we conducted pilot tests, comparing it to a \emph{Wizard-of-Oz} method. In this scenario, a human \enquote{simulated a \ac{CNN}} to explore user interaction with such a system.

\change{

\subsubsection{Point of Time to Communicate the Suggestion}
\label{sec:admc-concept-threshold}
Our \ac{ADMC} concept uses an adaptive \ac{DoF} mapping system to recommend \ac{DoF} mappings to the users depending on the current situation. The system visualizes the currently active \ac{DoF} mapping as a bright cyan and the suggestion as a dark blue arrow (see~\autoref{fig:adaptive-suggestions}).
This suggestion can be communicated -- based on the the configuration -- either continuously or only if the next most likely movement direction differs from the currently active \ac{DoF} mapping by a certain threshold.
} 

To calculate this threshold -- the difference between the currently active and new most likely \ac{DoF} mapping --, \emph{cosine similarity}~\cite{singhal2001modern} is used, ranging from exact alignment [0\%] to total opposite direction [100\%]. The formula for cosine similarity of two n-dimensional vectors is defined as: 

\begin{equation}
\text{cosine similarity} = \cos (\va*{a}, \va*{b}) = \frac{\va*{a} \va*{b}}{\|\va*{a}\| \|\va*{b}\|} = \frac{ \sum_{i=1}^{n}{a_i b_i} }{ \sqrt{\sum_{i=1}^{n}{(a_i)^2}} \sqrt{\sum_{i=1}^{n}{(b_i)^2}} }
\end{equation}

To implement a difference value, the cosine similarity needs to be transformed. As a cosine similarity of -1 indicates completely opposed vectors, the difference value needs to return 1 -- i.e. the maximum possible difference -- for a cosine similarity value of -1. A cosine similarity of 1, indicating exact similarity, should return a difference value of 0 -- i.e. no difference. Perpendicular vectors with cosine similarity 0 should return a difference value of 0.5 -- i.e. a 50\% difference. To calculate the difference value \textbf{d}, the following formula is used:

\begin{equation}
\text{difference d}  = 1 - \frac{\cos (\va*{a}, \va*{b}) + 1}{2}
\end{equation}

This difference value represents the difference between two vectors. While the user moves the robot with an active \ac{DoF} mapping, the adaptive \ac{DoF} mapping system reevaluates the situation and calculates new suggested \ac{DoF} mappings. The default difference value is set to 0.2 (20\% difference between currently active and new most likely \ac{DoF} mapping).

\change{

\subsection{Full Mixed Reality Continuum}
\label{subsec:MRcontinuum}

In our framework, we created an environment in which the entire continuum of \ac{MR} is exploitable. This extends the use of \emph{AdaptiX} to new scenarios and environments -- including the real world. 
The virtual and real environments of the robotic arm are aligned, allowing researchers to seamlessly switch between the user controlling the real and virtual robot. The level of \ac{MR} can be adjusted in various steps (cf. the \emph{virtuality continuum} of \citet{milgram1994taxonomy}). 

The \ac{MR} environment setups include:
\begin{enumerate}
    \item the completely real environment with the real robotic arm,
    \item the real environment extended with visual cues,
    \item the real environment into which the virtual robot is transferred and displayed (with and without visual cues),
    \item the virtual environment into which the real robot is transferred and displayed (with and without visual cues),
    \item the completely virtual environment with the virtual robotic arm.
\end{enumerate}

\begin{figure}[tbp]
\centering
    \subfloat[]{\includegraphics[width=0.325\linewidth]{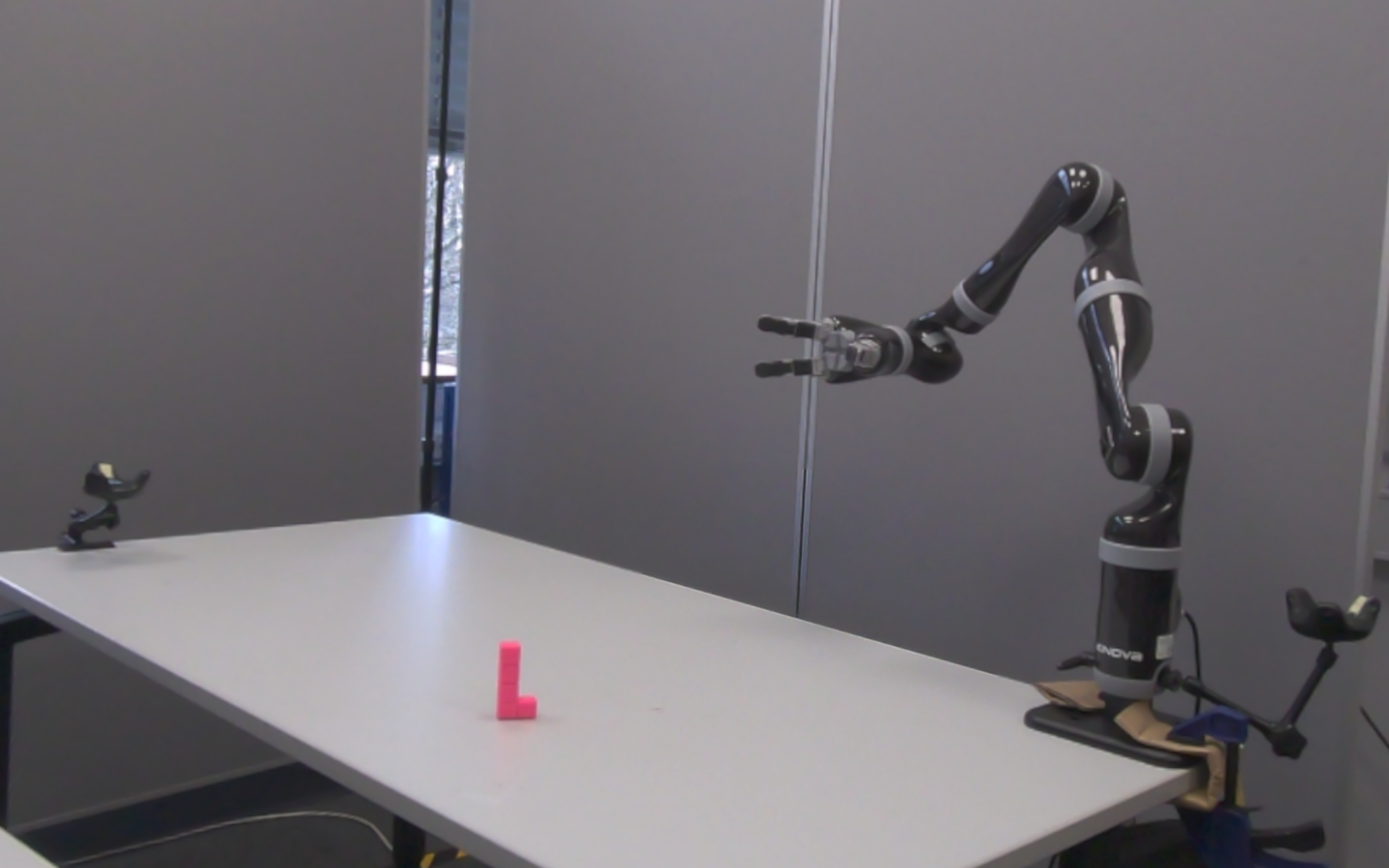}
    \label{fig:mr-continuum_a}}
    \hfill
    \subfloat[]{\includegraphics[width=0.325\linewidth]{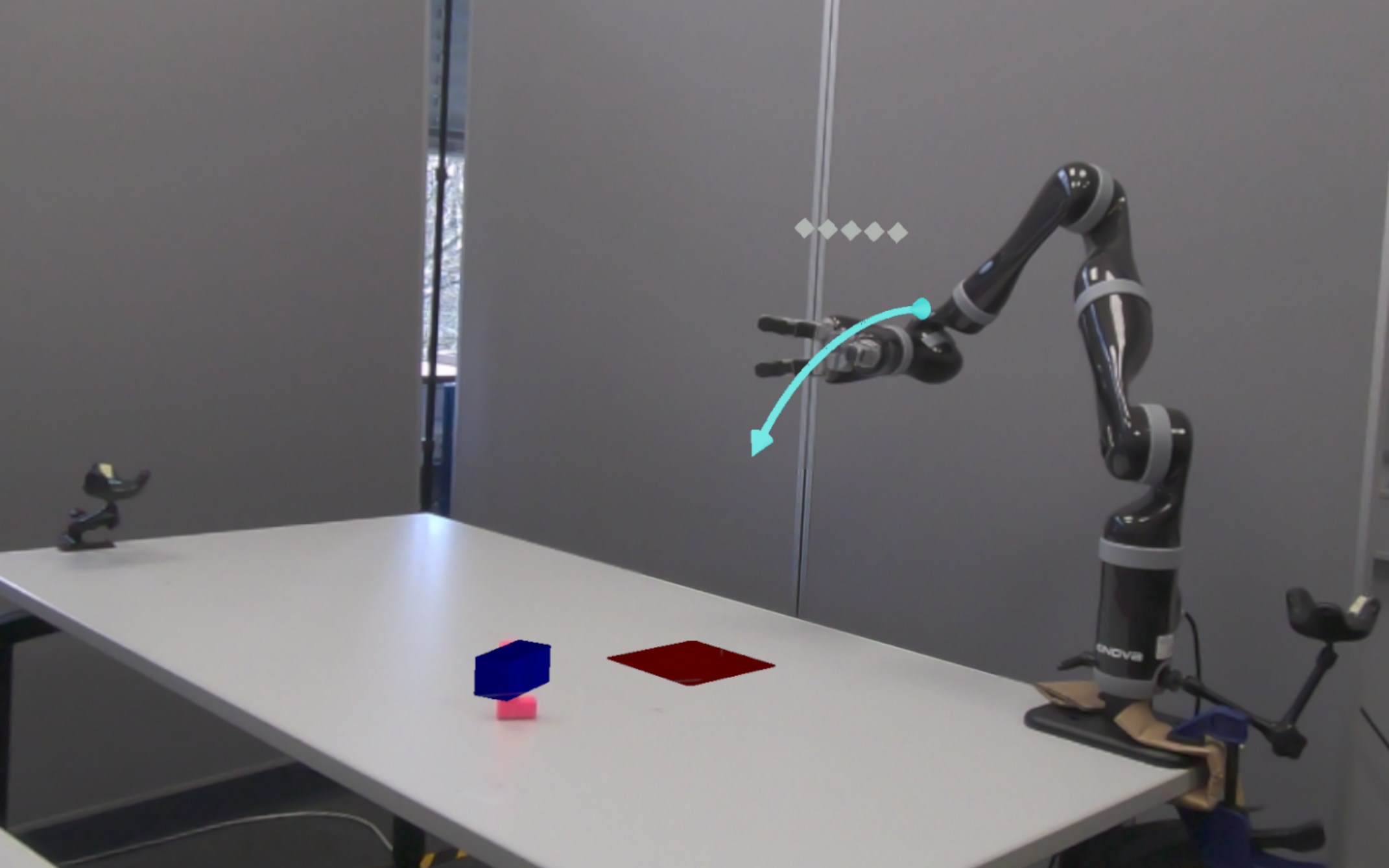}
    \label{fig:mr-continuum_b}}
    \hfill
    \subfloat[]{\includegraphics[width=0.325\linewidth]{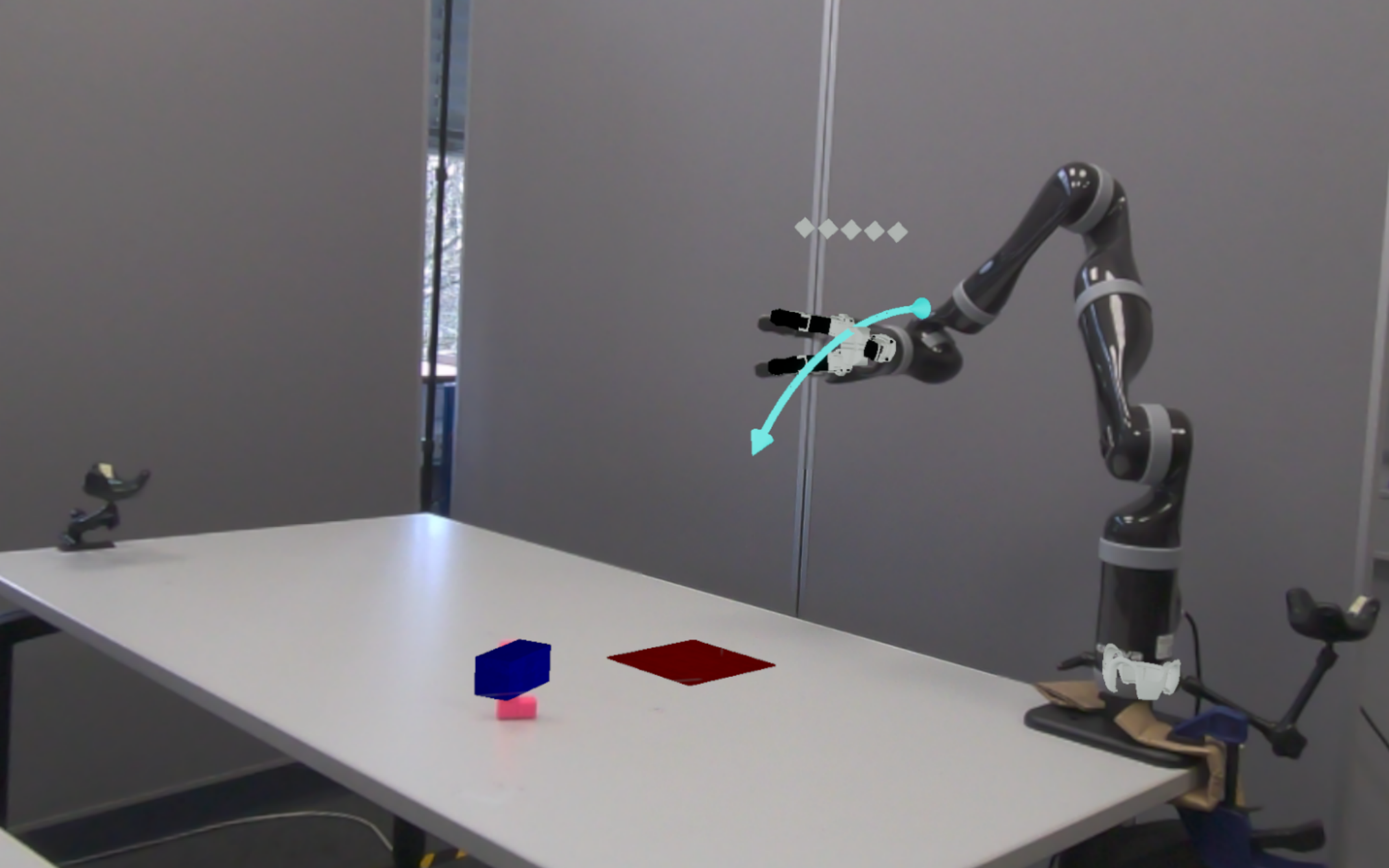}
    \label{fig:mr-continuum_c}}
    \hfill
    \subfloat[]{\includegraphics[width=0.325\linewidth]{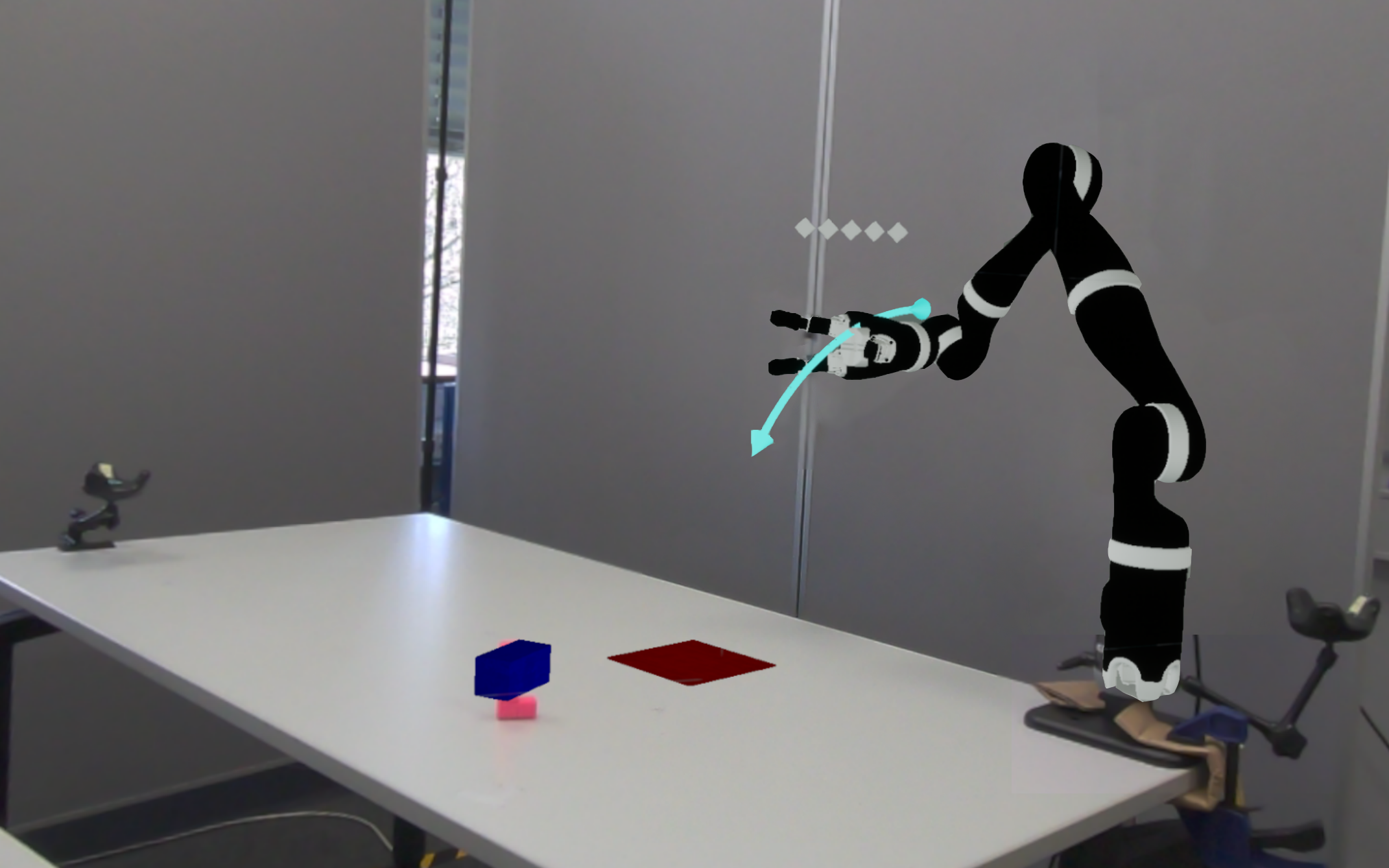}
    \label{fig:mr-continuum_d}}
    \hfill
    \subfloat[]{\includegraphics[width=0.325\linewidth]{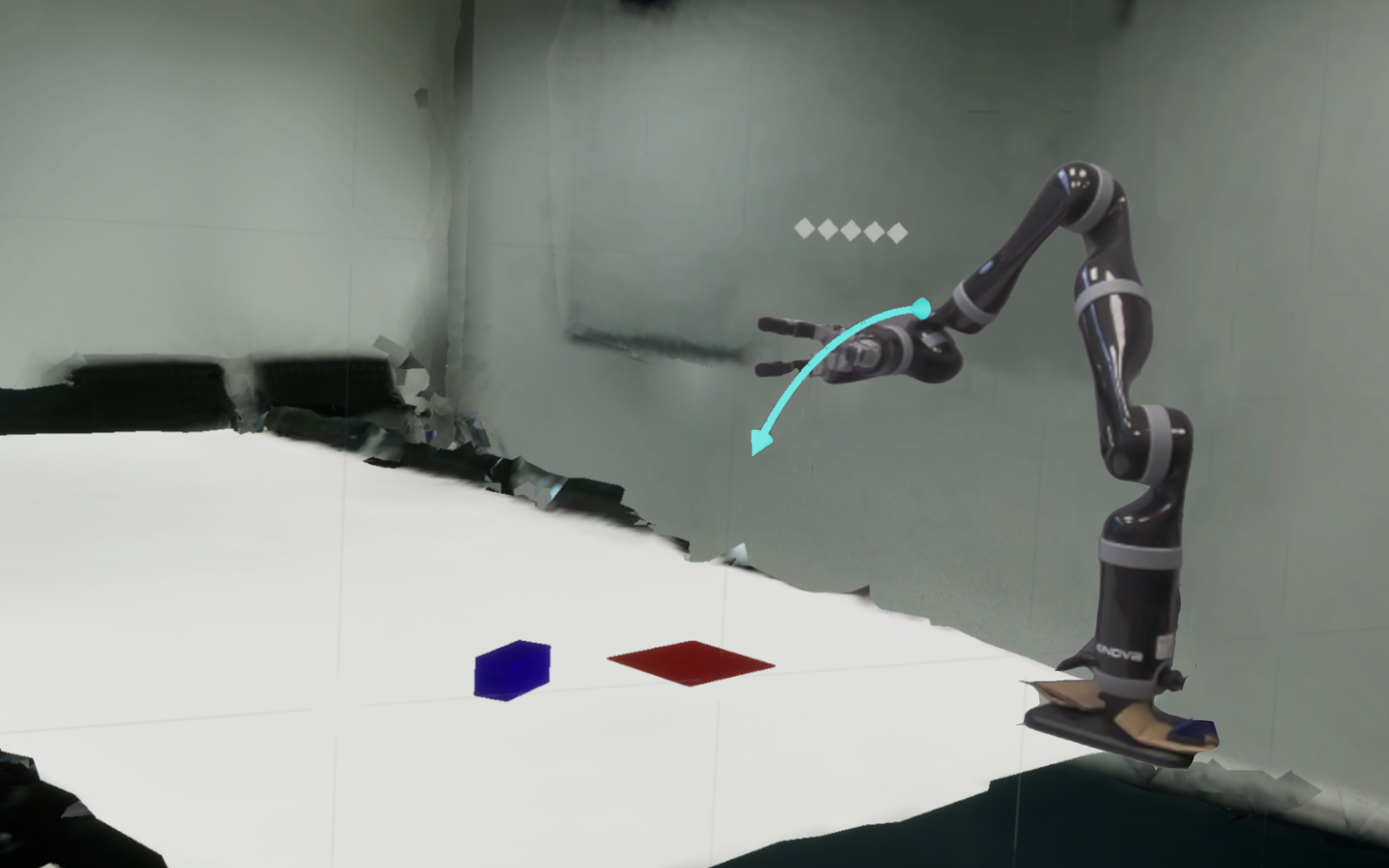}
    \label{fig:mr-continuum_e}}
    \hfill
    \subfloat[]{\includegraphics[width=0.325\linewidth]{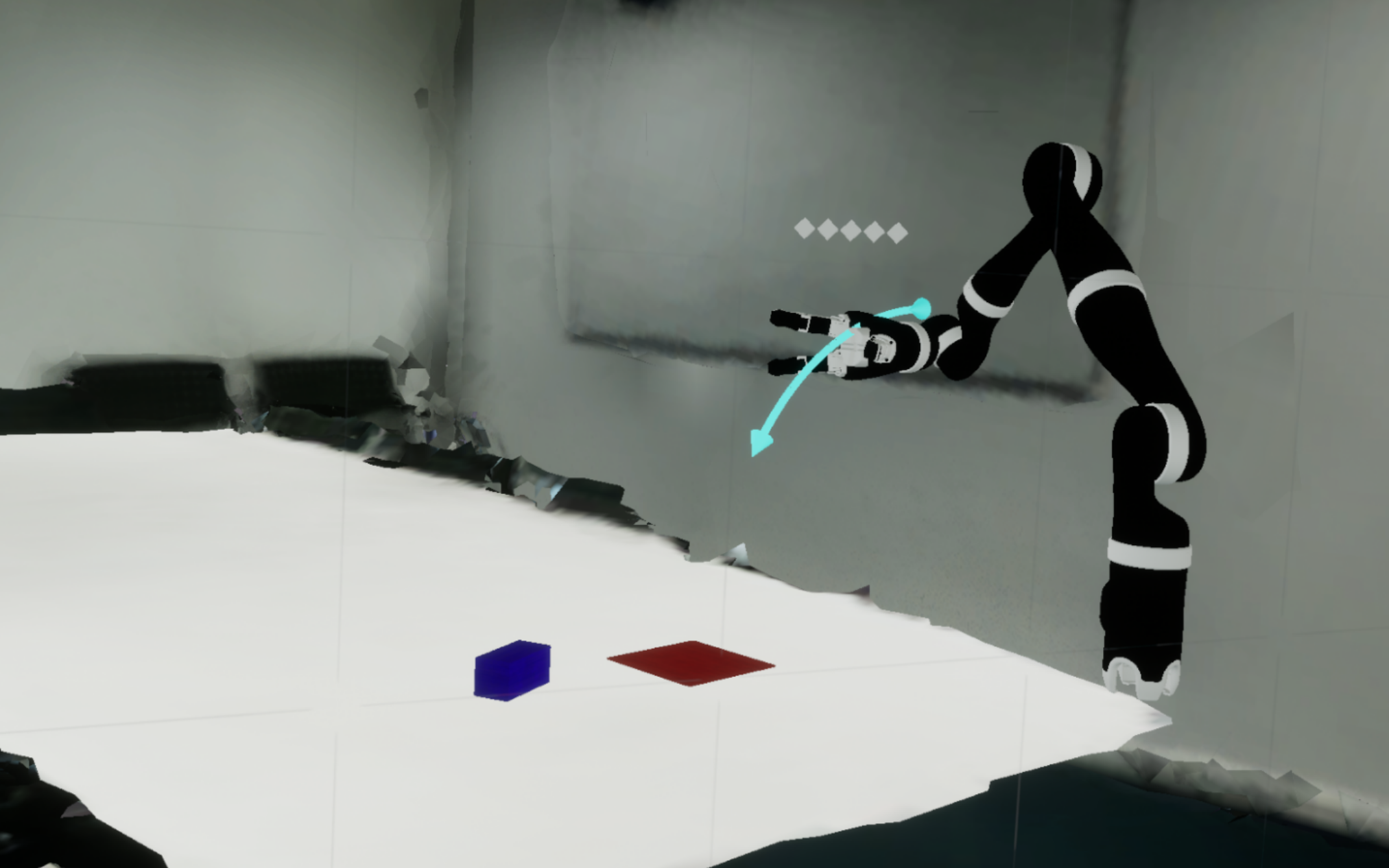}
    \label{fig:mr-continuum_f}}
\caption{\ac{MR} continuum with (\textbf{a}) only the real robotic arm in real environment, (\textbf{b}) augmenting of directional cues in the real environment with the real robotic arm, (\textbf{c}) additional visualizing the gripper and base of the virtual robotic arm in the real environment, (\textbf{d}) visualizing the simulated robotic arm in the real environment, (\textbf{e}) visualizing the real robotic arm in the virtual environment, and (\textbf{f}) visualizing the simulated robotic arm in the virtual environment.}
\label{fig:visualizations-continuum-mr}
\end{figure}

A comparison of the user's view in reality and simulation can be seen in \autoref{fig:visualizations-continuum-mr}.
\ac{MR} continuum level (1) is suitable for study baseline-condition, without any multi-modal feedback to the user. 
In level (2) an \ac{AR} visualization technique is mimicked, showing the whole physical setup augmented by basic cues.
Especially level (3) and (4) enable customizing either the robot itself or the environment to extent/exchange the physical setup but still not loosing the context. In (3) users can interact with a totally new or customized robot while being in a familiar environment. World's distractions can be excluded in (4) while the the original robot is presented.
Finally, level (5) provides a \ac{VR} environment that can be fully customized.

} 

\change{
\subsection{Interfaces}
\label{sec:fw-interfaces}
} 
We designed \emph{AdaptiX} to facilitate the comparison of different interaction designs, intervention strategies, and feedback techniques for shared robot control. The initial version of the framework includes interface types for extending user input, \ac{ROS} integration, and multi-modal feedback. However, this baseline can easily be customized and extended by future development.

\subsubsection{User Input}
\label{sec:fw-userInput}
We provide a standard control approach where pressing a keyboard button moves the end effector along cardinal \acp{DoF} (x, y, z, roll, pitch, yaw, opening and closing the gripper). Using build-in functionalities, the designated keyboard input can easily be adjusted to other input devices like gamepads, joysticks, or customized assistive input appliances.

In contrast to tele-operating the robotic arm, a \emph{follow-me} approach for any trackable object in 3D space -- e.g., the user's handheld \ac{VR} motion controller -- was implemented. 
The robot's end effector directly follows the movement of the trackable object, which corresponds functionally to direct control. This can be used to generate high-dimensional input and record intended behavior quickly, providing an easy way of interacting and controlling the robot, especially for inexperienced users.

\subsubsection{ROS Integration}
\label{sec:fw-ros}
\begin{figure}[htbp]
\centering
\includegraphics[width=\linewidth]{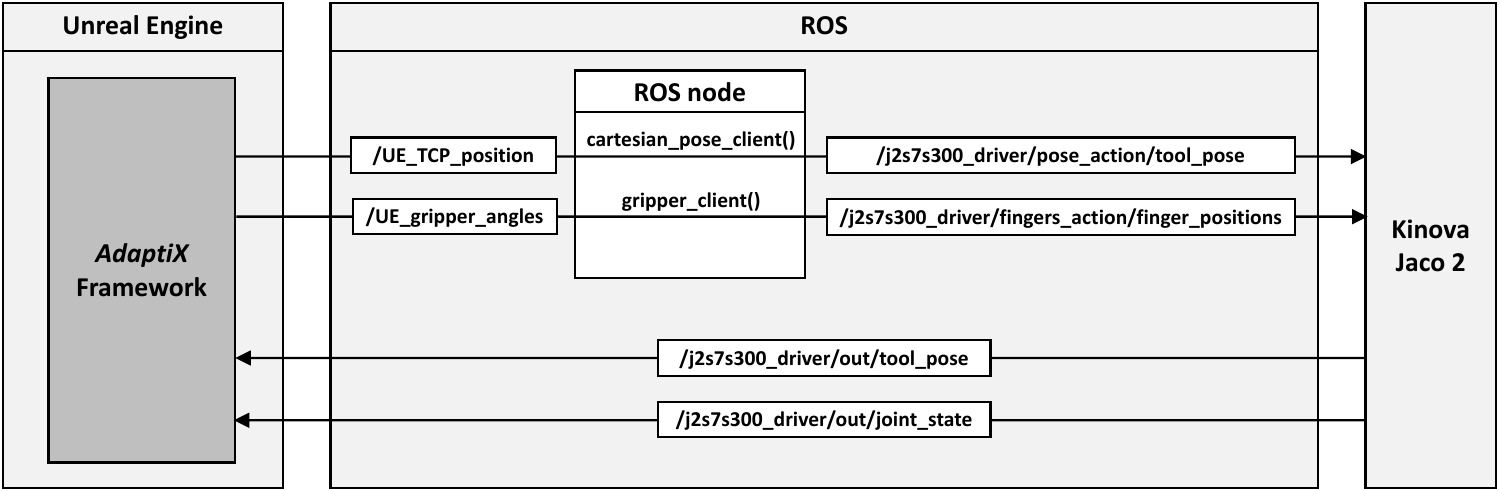}
\centering
\caption{\change{Component connections of the \ac{ROS} interface for mixed reality.}}
\label{fig:ros-interface}
\end{figure}

The \ac{ROS} integration allows for a bidirectional exchange of information between the simulation and a real robot, mirroring the robot's state \emph{in-silico} and vice versa. 
\autoref{fig:ros-interface} shows the involved components: a \ac{ROS} bridge facilitates the multi-device connection between the framework and the real robot while exchanging robot data. 
On the \ac{ROS} side, the messages for the arm position and orientation control and the values for the angle-accurate control of the gripper fingers are read in via the \ac{ROS} subscriber node. They are then processed, and the robot arm and gripper are controlled through our action client.
In addition, the joint angles, the \ac{TCP}, and the position of all three gripper fingers are published via \ac{ROS}, which are then input by our \emph{Unreal Engine} framework. The virtual and real robots are synchronized via \ac{ROS} every 0.1 seconds.

Based on this, our framework provides -- depending on the specific context -- both a \emph{DigitalTwin} and \emph{PhysicalTwin} approach, allowing the control of either with the other. 

\change{
\subsubsection{Multi-Modal Feedback}
\label{sec:fw-multiModalFeedback}
To communicate any combination of \acp{DoF}}, our framework supports several visual cues to illustrate the intended movement trajectory and provides multi-modal feedback extensions via audio and haptic-tactile feedback. Visual feedback can be either provided dynamically attached to the virtual/physical robot's end effector, stationary in the world, or attached to the user's view.

\emph{AdaptiX} aims to support the development of novel multi-modal interaction and feedback designs either in the pure \ac{VR} simulation testbed environment or by interacting with a real robot in \ac{MR}, which mimics an \ac{AR} setting due to the stereoscopic video-feed. 
Moreover, it \change{is} also possible to show the real robot in our \ac{VR} simulation environment instead of the simulated one.

\begin{figure}[htbp]
\centering
\subfloat[Ghost]{\includegraphics[width=0.32\linewidth]{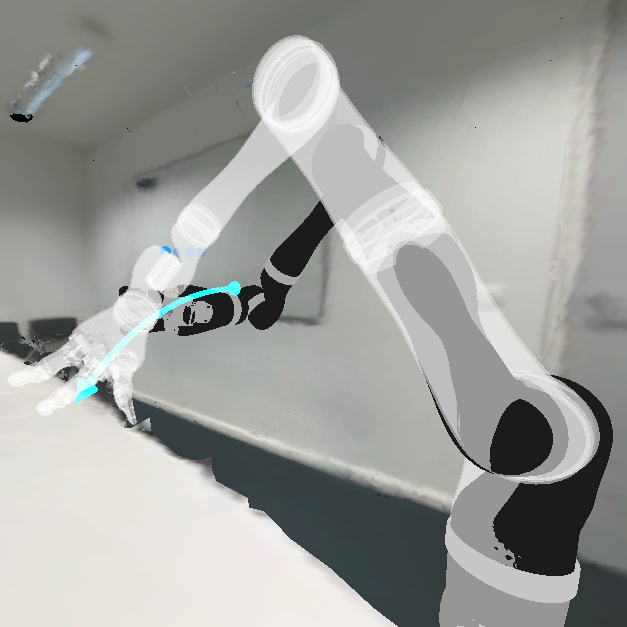}}
  \hfill
  \subfloat[Waypoints]{\includegraphics[width=0.32\linewidth]{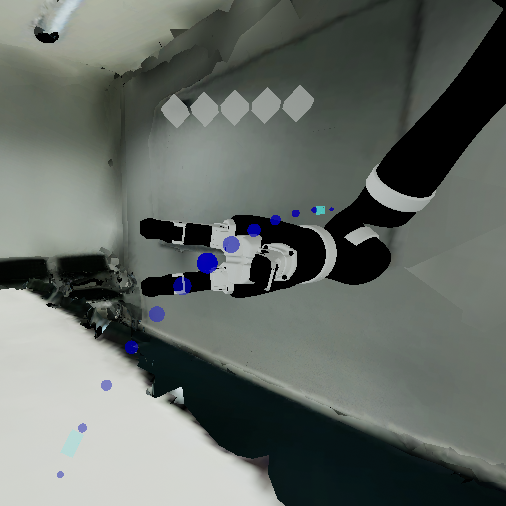}}
  \hfill
  \subfloat[Arrows]{\includegraphics[width=0.32\linewidth]{figures/adaptive_suggestions_0.png}}
\centering
\caption{Visualization examples pre-implemented in the framework.}
\label{fig:visualizations-interfaces}
\end{figure}

\autoref{fig:visualizations-interfaces} shows three exemplary \ac{AR}-style visualizations provided by the framework, including \change{(a)} a robotic ghost overlay, \change{(b)} discrete waypoints in 3D, and \change{(c)} a variety of multidimensional arrows. Though varying in design, these visualizations can effectively communicate the robot's motion intent to the user.

\textbf{Ghost:} A visualization of robot motion intent by showing an additional version of the robot (or specific components) registered in 3D space, in another color and/or opacity. These visualizations communicate the exact position and orientation a robot at a given time, behaving precisely as though the real robot had been moved this way.

\textbf{Waypoints:} This visualization technique augments the position of a robot (or in our case, the gripper of the robotic arm) in 3D space at a certain point in the future. Usually, the robot navigates linearly between these \emph{Waypoints}, which increases predictability.

\textbf{Arrow:} Among visualizations arguably the most basic but certainly also the most familiar (as seen in traffic navigation systems, road signs, and on keyboards). \emph{Arrows} are found both in straight and curved varieties, where curved arrows indicate a rotation. Given the abundance of \emph{Arrows} in daily life, it makes sense that many robot motion intent visualizations use them.

\textbf{Classic:} This visualization also uses \emph{Arrows}, but in our prototype they are used as a baseline condition to evaluate adaptive and non-adaptive controls. Here, as with the standard input device \emph{Kinova Jaco 2}, two axes can be controlled simultaneously and the user has to choose between different translations and rotations by mode-switching.

All interfaces are modular, enabling quick adaptations and switching between variations. This flexibility allows for studies with clean methodologies and easy comparisons without additional overhead. 
The community is invited to extend the implementations with any interfaces or control methods desired for their research.
\subsection{Recording and Replay}\label{sec:fw-recording}

\emph{AdaptiX} contains an easy-to-use general-purpose system to record, store and replay simulation data, including detailed information about robot states, execution times, or the states of various objects in the environment. The recording system generates \ac{CSV} text files, which can be accessed with any data manipulation software (e.g., Python or MATLAB). The added output functionality differs significantly from the replaying system provided by the underlying \emph{Unreal Engine}, which is mainly designed for visual replays and -- among other things -- does not support a \ac{CSV} file format.

In addition, \emph{AdaptiX}'s recording and replaying system is entirely customizable. Camera re-positioning or re-rendering background scene options are included in the initial version. By default, the recording system tracks the user's view, the robotic arm, and all moveable actors in the virtual environment. All other objects are assumed to be stationary, thus part of the level, and ignored as such. This approach allows for the randomization of background scenes by re-rendering. 

The system stores the assigned virtual meshes, scales, possible segmentation tags for each tracked object, and the complete pose data per frame. 
During the replay process, all objects that were initially recorded in a specific level are swapped with the corresponding data stored in the loaded recording. However, if a different scene is being loaded, the objects from that scene are used instead.
In every subsequent frame, all objects are positioned at their respective position until the loaded recording has finished. The system permits custom code to be run at the end of each loaded frame, thus enabling de-bugging and data rendering during replays. 

Overall, \emph{AdaptiX} facilitates the lightweight storage of recordings as \ac{CSV} files with the option to render and store complex and large-scale data (e.g., images or videos) for subsequent evaluation. This lightweight approach is particularly useful when deploying experiments on external devices or recording extensive datasets.
\change{
\section{Framework Implementation}
\label{sec:framework-overview}
}
The \emph{AdaptiX} simulation environment is based on the game engine \emph{Unreal Engine 4.27}~\cite{unrealengine}. The advanced real-time 3D photoreal visuals and immersive experiences provide a suitable foundation for our framework, and assets for future extensions are readily available. \emph{Unreal Engine 4.27} includes integrated options for various hardware setups, thus enabling the framework to be deployed on different operating systems while utilizing most currently available \ac{VR}/\ac{MR}/\ac{AR} headsets, gamepads, and joysticks. 
At the time of writing, \emph{Unreal Engine 4.27} is free to use, has a considerable user space, and allows unrestricted publications of non-revenue generating research products like the \emph{AdaptiX} framework. \change{Detailed implementation descriptions can be accessed in the \emph{README} provided in the repository at \url{https://adaptix.robot-research.de}.}

\begin{figure}[htbp]
    \centering
    \includegraphics[width=\columnwidth]{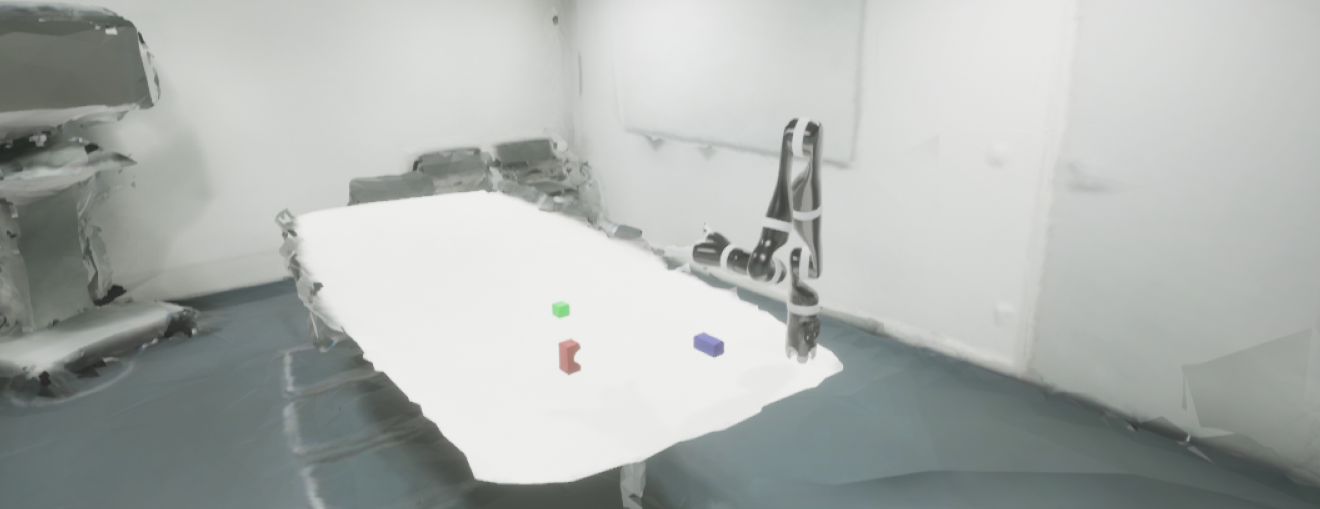}
    \caption{Example scenario provided in \emph{AdaptiX} including a table, a virtual \emph{Kinova Jaco 2} robotic arm and colored blocks on the tabletop.}
    \label{fig:sim_envr}
\end{figure}

\change{
\subsection{Simulation Environment}
\label{sec:sim-env}

The \emph{AdaptiX} default scenario centers on the photogrammetry scan of an actual room that contains a table with an attached virtual robotic arm (see \autoref{fig:sim_envr}). A simulated camera is mounted on the arm's gripper. We added a toggle-off option to hide the camera from the user's view.
}

The framework includes a straightforward testbed scenario for pick-and-place operations, mimicking the basic principles of most \ac{ADLs}. The simulation centers around a red surface as a drop target and a blue block as the to-be-manipulated object. Once the object has been successfully placed, the setup randomly re-positions the blue block on the table surface, and the task can be repeated.

We optimized the robotic arm simulation for operation via a \ac{VR} motion controller with an analog stick, several playable buttons, and motion capture capabilities (e.g., \emph{Meta Quest 2}~\cite{metaquest2}). These options provide a workable foundation to implement and test diverse interaction concepts, including adaptive concepts which can be configured to match the individual physical abilities of the intended user. 

\change{
By incorporating the \emph{Varjo XR-3}~\cite{varjo} -- a particularly high-resolution \acs{XR}-\ac{HMD} -- we implemented a transitional \ac{MR} environment. Using two \emph{HTC VIVE} trackers~\cite{vivetracker}, the virtual and real worlds are synchronized so that the robots' working areas are identical. By including the \emph{HTC VIVE} motion controller~\cite{vivecontroller}, it is then possible to control the physical robot directly via the \emph{PhysicalTwin} approach of \emph{AdaptiX} (see~\autoref{fig:teaser}). 
}

The virtual robotic arm is designed as a modular entity, allowing easy integration to new levels following the \emph{Unreal Engine}'s \emph{ActorBlueprint} class structure.

\subsubsection{Simulated Robotic Arm}
\label{sec:robot-arm}

The commercially available \emph{Kinova Jaco 2} assistive robotic arm~\cite{kinova} is specifically designed as an assistive device for people with motor impairments. It is frequently used by a) the target audience and b) researchers -- e.g.,~\cite{Beaudoin.2018,Herlant.2016modeswitch} -- during \ac{HRI} studies, hence the suitability for inclusion in \emph{AdaptiX}. 

\begin{figure}[bp]
\centering
\includegraphics[width=\columnwidth]{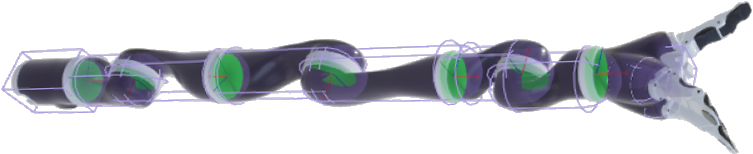}
\centering
\caption{Virtual Robotic Arm with Physics Constraints: purple capsules represent links, green discs represent angular constraints.}
\label{fig:jaco-physics}
\end{figure}

We designed the simulated \emph{Kinova Jaco 2} as close as possible to the actual product, using virtual meshes generated directly from \ac{CAD} files provided by the manufacturer. Much like in reality, the virtual arm consists of a series of individual links connected by angular joints as shown in the annotated rendering of the assembled model \autoref{fig:jaco-physics}.  

As \emph{AdaptiX} -- including the operation of its simulated robotic arm -- is optimized for \ac{HRI} studies, it focuses on user interaction rather than low-level robot control, whilst also able to incorporate those. Hence, rather than following the standard base-up control, the simulated arm moves in reverse: the user's input directly controls the end effector's motion; the connected joints are positioned to connect the end effector with the base.Each intermediate joint is modeled as a dampened spring with the links unaffected by gravity. This also resolves the redundancy, i.e., joint angle ambiguity a 7-jointed robot has.

This approach allows for nearly arbitrary motion of the end effector and a semi-realistic interaction of the arm with the environment. As a beneficial side effect, developers can disconnect the end effector from the rest of the arm and allow the user to control a free-floating robot hand without any constraints. However, the internal physics engine to realistically handle collisions and interactions between the end effector and the environment \change{is still active}.

Likewise, \change{w}e based the grasp concept on a custom interaction design for robotic grasping rather than physics. Physics-based grasping in a virtual environment is a challenging task~\cite{physics_grasp} and would require substantial preparation and asset fine-tuning from future developers who use the framework. Instead, we defined a logic-based approach that we consider sufficiently realistic for shared control applications: an object is regarded as grasped once it has contact with two opposite fingers while closing the gripper until the fingers open again. The grasped object is rigidly attached to the end effector, keeping its relative position stable and moving alongside the end effector until released.

\subsubsection{Simulated Camera System}
Computer-aided robot control usually requires a camera system -- or a comparable sensor -- to measure context information about the current environment for the underlying software function. To provide a realistic equivalent in simulation, \emph{AdaptiX} contains a virtual version of the commercially available \emph{Intel Realsense D435}~\cite{D435_Datasheet}. This camera system is commonly used in research applications~\cite{Ciccarelli.2022.realsense,Zhang2019.realsense} and can deliver aligned color and depth images. 
The built-in color sensor generates depth data by applying a stereo-vision algorithm using grayscale image data of two built-in \ac{IR} imagers. To improve the texture information captured by the \ac{IR} imagers, the camera also includes an \ac{IR} projector, which projects a static pattern on the scene.

As with the simulated robotic arm, the virtual camera system is a modular actor that can be arbitrarily placed within the simulation environment. Its mesh and texture are derived directly from the manufacturer's \ac{CAD} files to optimize authenticity. The virtual camera system includes all image sensors of the original, plus an optional virtual sensor generating a segmented image of the scene. We designed the virtual sensor parameters to be as close as possible to those of the actual sensors. They include -- but are not limited to -- sensor dimensions, lens structure, focal length, and aperture.

Because the framework can provide depth information directly from the 3D simulation, the virtual depth camera does not need to calculate its data using stereo-vision but instead yields perfect per-pixel depth information. If stereo-vision-generated depth data with realistic noise, errors, and other algorithm-specific effects is needed, the virtual system also delivers the \ac{IR} images for a manual calculation.

Additionally, the simulated camera system supports the usage of the image data in-simulation and storing the data on disk for applications such as dataset generation or logging.
\subsection{Adaptive DoF Mapping Control (ADMC)}

The adaptive \ac{DoF} mapping is implemented in the object \emph{Axis Wizard}, which provides functions to calculate the optimal suggestion, as well as the other possible optimizations. 
The calculation relies solely on the virtual objects in the simulation environment instead of object recognition or camera data to enable development and evaluation without a physical robot setup.
However, the camera feed for object recognition can be activated by developers to read positions and orientations. 
In addition to the positions and orientations of the \emph{Gripper Mover} and the \emph{Current Target} (which can be an object to pick up or a target surface to place the object on, depending on the context), two other parameters of \emph{Axis Wizard} are important to ensure the correct calculations for the pick-and-place task -- \emph{Minimal Hover Distance} and \emph{Hover Height}. 

Disregarding the handling of edge cases, the calculation of the optimal suggestion is taken care of in three steps: 1) calculating \emph{Translation}, 2) calculating \emph{Rotation}, and 3) calculating the finger movement variable \emph{Gripper}.
\change{The Blueprints for implementation details are provided in \autoref{sec:appendix}.}

\subsubsection{Calculation of the Optimal Suggestion}

\emph{Minimal Hover Distance} represents the distance -- projected on the XY-plane -- between the \emph{Gripper Mover} and the \emph{Current Target}. When this distance is smaller than the \emph{Minimal Hover Distance} (see \autoref{fig:axis-1-translation} \change{in the appendix}), the \emph{Axis Wizard} uses a point above the \emph{Current Target} for its calculations -- referred to as the \emph{Target Point}, instead of the \emph{Current Target}'s position to prevent the robot from getting too close to the table, allowing for proper gripper rotation.
Then, a vector from the \emph{Gripper Mover}'s position towards the \emph{Target Point} is calculated, normalized, and inversely rotated by the \emph{Gripper Mover}'s rotation. This calculation returns a unit vector pointing from the \emph{Gripper Mover} toward the \emph{target point} in the \emph{Gripper Mover}'s reference frame.
This vector is then scaled by the \emph{Vel Trans} value of the \emph{Kinova Jaco 2} to get a translation of the size of the movement performed by the \emph{Kinova Jaco 2} during one frame. 

\emph{Hover Height} determines the height of the aforementioned point above the \emph{Current Target}. If the XY-projected distance between the \emph{Gripper Mover} and the \emph{Current Target} is smaller than the \emph{Minimal Hover Distance}, the \emph{Axis Wizard} directly uses the \emph{Current Target}'s position for its calculations instead of the point above it.

To calculate the optimal suggestion's \emph{Rotation}, the \emph{Translation} -- calculated in the first step -- is used as input for the \emph{Make Rot from X} node. 
This node returns a \emph{rotator} representing the rotation required to make an object point toward the direction indicated by the input vector -- \emph{target point}.  
To mitigate an additional \emph{roll} of \emph{Gripper Mover}, the inverse value is added, keeping the \emph{Gripper Mover's} orientation largely steady. 
Additionally, since only a small part of the rotation is performed during one frame, the \emph{rotator} is scaled down. The calculation for the \emph{Rotation}, excluding edge cases, is depicted in \autoref{fig:axis-1-rotation} \change{in the appendix}.

\subsubsection{\change{Calculation of Gripper values}}
The \emph{Gripper} value only depends on whether the target point is within reach of the robotic fingers, either with or without additional movement (i.e. if the fingers are almost close enough, there will be a movement towards the target point, otherwise the fingers will engage without moving the gripper) and whether or not an object is currently being grasped (i.e. if an object is grasped and the gripper is close to the target point, it suggests to open the fingers, otherwise close them).

\subsubsection{Calculation of the Adjustment Suggestion}

The adjustment suggestion is calculated by rotating the optimal suggestion's \emph{Translation} by 90° around the Y-Axis, keeping the same \emph{Rotation} and setting the \emph{Gripper} value to 0.  
This results in a \ac{DoF} mapping which moves roughly along the \emph{Gripper Mover's} Z-Axis, or colloquially "up and down" between the fingers if the optimal suggestion is seen as "forward and backward". As \emph{Rotation} is kept the same between the optimal and adjustment suggestions, the resulting movement keeps the fingers roughly facing the direction of the \emph{Current Target}. 

The translation, rotation, and gripper suggestions use much simpler calculations. The translation suggestion calculates a vector from the \emph{Gripper Mover} towards the \emph{Current Target}, inversely rotates it by the \emph{Gripper Mover's} rotation to put it into the \emph{Gripper Mover's} reference frame and uses that as the \emph{Translation} value for the suggested \emph{Adaptive Axis}. This vector is also what the rotation suggestion uses to calculate a \emph{Rotator} representing a rotation towards the \emph{Current Target}. The gripper suggestion checks whether an object is currently being grasped. If so, the suggestion is to open the fingers (\emph{Gripper} = -1). Otherwise, the suggestion is to close the fingers (\emph{Gripper} = 1). 

\subsubsection{Attention Guidance in \emph{Threshold}}

Both the \emph{Continuous} and \emph{Threshold} approaches share the same core calculation for \ac{DoF} mappings. However, the \emph{Threshold} approach has an additional task: determining whether the optimal suggestion significantly differs from the currently active \ac{DoF} mapping. This task is more related to visualization than the \ac{DoF} mapping calculation itself and is managed by the \emph{Gizmo} object.

The \emph{Gizmo} object contains a \emph{Realtime Threshold} variable, which represents the threshold as a value between 0 and 1. It also includes a function called \emph{Adaptive Axes Nearly Equal}, which determines whether two \emph{Adaptive Axes} are nearly equal by checking if their difference is below the \emph{Realtime Threshold}. The threshold value is chosen to be between 0 and 1 to align with a percentage of difference (see \autoref{sec:admc-concept-threshold}), providing a more intuitive understanding of the amount of difference compared to the cosine similarity value used as the basis for the difference calculation.

As the \emph{Unreal Engine} does not provide an arbitrarily sized vector structure, the calculations required needed to be programmed manually rather than with built-in vector operations. Therefore, two math expression nodes were defined, one calculating the dot product of two 7D vectors and the other calculating the magnitude of a 7D vector. Using these, the cosine similarity between two \emph{Adaptive Axes} could be calculated in \emph{Unreal Blueprints} (see \autoref{fig:adaptive-axes-nearly-equal} \change{in the appendix}). To forego the transformation of the cosine similarity into a percentage difference, the \emph{Unreal Engine's Nearly Equal} node was used to determine whether the cosine similarity was nearly equal to \emph{1} -- meaning the vectors align -- with a threshold of \emph{2 * Realtime Threshold}. The threshold needed to be multiplied by \emph{2} as the range of the cosine similarity has a magnitude of 2. The result of this calculation is a boolean value that is true if the difference between the \emph{Adaptive Axes} is below the threshold and false otherwise.

The resulting value is then used by the \emph{Gizmo} to show the arrow corresponding to the optimal suggestion. It is also used to notify the \emph{Game Mode} -- an object representing the game, keeping track of study variables, etc. -- that the threshold was broken. This triggers an event that causes a 1kHz sound to play and a haptic effect to occur on the motion controller. A reset variable is used to prevent the sound from constantly triggering. However, there appears to be a specific point during movement at which it is possible for users to stop their input and the software to get caught in a loop of firing the event and resetting it, causing a constant sound and vibration. If users continued their movement, the software stopped firing the event, seizing the sound and vibration. Unfortunately, this was only noticed during the experiment, which is why the problem persists in the current software version. Assuming \emph{Threshold} is to be used in future research, a better solution for a single fire execution of the notification needs to be developed.

\section{Limitations}
\label{sec:limitations}
In \ac{HRI} research, the leading factor impacting user experience is usually the chosen method of (shared) control and the respective interfaces. Using frameworks like \emph{AdaptiX} allows researchers to tweak these variables toward high user satisfaction through methodological studies and experiments.

However, like any simulation, \emph{AdaptiX} only approximates reality and contains ingrained limitations when working with the system and evaluating generated results.

\subsection{Scenario Selection}
In the initial version, \emph{AdaptiX} provides only a single level, as seen in all screenshots of this work. This scenario functions mainly as a model for simple tasks. As such, it lacks environment interactions or varying backgrounds and is not designed for a specific assistive task.

This single level might need to be revised to represent the complete application range of assistive shared control, which is why extensions are required. As such, \emph{AdaptiX}'s modular design allows the community to generate custom levels for their specific research interests effortlessly. 

\subsection{Simulation Sickness caused by Head Mounted Display}

\acp{HMD} are a popular tool to create immersive virtual environments, frequently used in research and industrial settings. However, using a \ac{HMD} in \ac{HRI} can create a significant displacement between the virtual object and the physical world through effects related to the resulting limited field of view, reduced depth perception, and distorted spatial cues.

For applications within the \emph{AdaptiX} framework, these issues could result in users experiencing motion sickness, disorientation, discomfort, and potentially decreased performance when interacting with the simulated robotic arm or virtual objects. Researchers must consider these artifacts when designing experiments, especially when developing studies including qualitative questionnaires or when comparing different levels of \ac{MR} continuum.

\change{
\subsection{Simulation Environment}
The simulation environment centers on the photogrammetry scan of an actual room that contains
a table with an attached virtual robotic arm. Compared to a 3D modeling of a room, the photogrammetry does not provide a high resolution, leading to a partial blurred appearance.

\emph{AdaptiX} does not provide a photo realistic virtual environment (yet). However, in our studies, the slightly blurred appearance never seemed to have had a negative effect. On the contrary, it has helped participants focus on the scene's relevant parts (i.e. the robot and objects). Researchers and developers are invited to create and evaluate a 3D modeled environment.
}

\subsection{Simulated Robotic Arm}

If controlled entirely in simulation, the robotic arm (as described in \autoref{sec:robot-arm}) does not move identically to an actual \emph{Kinova Jaco 2} because of implementation decisions favoring physical interactions over accurate per-joint robot actions. In most other cases, the individual joints are in relatively realistic positions, even though they might not be identical to the underlying solution provided by an inverse kinematic of the real robot. 

Especially in the \emph{follow-me} approach (see \autoref{sec:fw-userInput}), it is possible to reach outside of the mechanical range of the robotic arm. Due to the entirely physics-based connection, this results in partially disconnected joints. However, this is only an issue of visualizing the robotic arm in the simulation environment and does not affect the control or the \ac{TCP} data recording.

Likewise, grasping simulated objects is based on a custom implementation, and grabbed objects are firmly attached to the end effector. Care must be taken for objects that are -- in reality -- too heavy for the gripper, have slippery surfaces, or have mechanical dimensions that make the object unstable when held. Theoretically, this \enquote{ideal kind of grasping} allows the virtual robot to move any arbitrarily large and heavy object. To address this, we added the object tag \emph{Graspable} that allows developers to define permitted -- and by omission -- unpermitted objects.

\subsection{Simulated Camera System}

Although the simulated camera is based on manufacturer \ac{CAD} files, comparison tests failed to deliver completely identical data to the actual recording system. These variances stem from environmental differences between simulation and reality, as light or dust/other particles in the air will cause effects in the produced image. However, these effects can be added in post-production or -- if required -- activated in the framework. By default, the respective settings are disabled as they would primarily introduce noise that not every developer might want.

On a technical level, the images generated by the virtual system differ slightly in terms of data types. The virtual grayscale \ac{IR} images consist of three identical color channels instead of a single channel in reality. Also, the virtual \ac{IR} and color images include an additional fourth alpha channel, which is not used in our framework. 
The generated depth data format also differs, as the actual camera system generates images as 16-bit unsigned integer, and the simulation provides them as 16-bit signed floats. The depth data generated by the framework is pixel-perfect, which ignores various camera system effects that occur in reality by the calculation of depth using stereo-vision.

All these technical differences are addressed within the framework through data transformation and should not noticeably affect the output of \emph{AdaptiX}. However, researchers and developers should be aware of these adjustments for future developments and extension.

\subsection{ROS Interface}
The \ac{ROS} interface connects the virtual with a real robot, each with its own environmentally-determined set of limitations. This results in some logical inconsistencies while using the interface. The obvious velocity limitations of the real system result in delayed execution if reality is to follow the simulation. \change{Therefore, the maximum velocity of the virtual robotic arm is  set automatically to the physical characteristics after enabling \ac{ROS}.}
Also, as the virtual joints are not controlled by an \ac{IK} but instead based on physics, the interface sends only end effector poses to the real robot, omitting individual joint poses. This may result in differing robot configurations, with only the end effector point being aligned in some instances.

When sending pose data from the real robot to the virtual twin in simulation, most of these restrictions do not apply. The simulated robot can move arbitrarily fast, and its configuration aligns automatically with the real system. The only restriction is that, by default, no further information about the natural environment is available, resulting in a relatively empty virtual environment if relying purely on the \ac{ROS} interface.

When designing expansions, developers also must be aware that \ac{ROS} and \emph{Unreal Engine} differ in handedness. \ac{ROS} is based on a right-handed coordinate system, while the \emph{Unreal Engine} uses a left-handed approach. \emph{AdaptiX} internally does the necessary transformation for the robotic arm but will not automatically calculate this for other position and orientation data, e.g., obstacles.
However, researchers can mitigate this by applying the provided coordinate transformation methods of the robotic arm to any further object.
\change{\section{Framework Example Adaptions}}
\label{sec:case-studies}
The \emph{AdaptiX} framework has been successfully used \change{and adapted} in \change{three} case studies evaluating interaction concepts and multi-modal feedback with remote and laboratory-based focus groups. 

\subsection{\change{Example Adaption 1}: Adaptive Control of an Assistive Robot}

In an initial study~\cite{Kronhardt.2022adaptOrPerish}, the \emph{AdaptiX} framework was used to explore the proposed \ac{ADMC} control method with associated visual cues for various \ac{DoF} mappings. 

\begin{figure}[htbp]
\centering
\subfloat[Classic]{\includegraphics[width=0.32\linewidth]{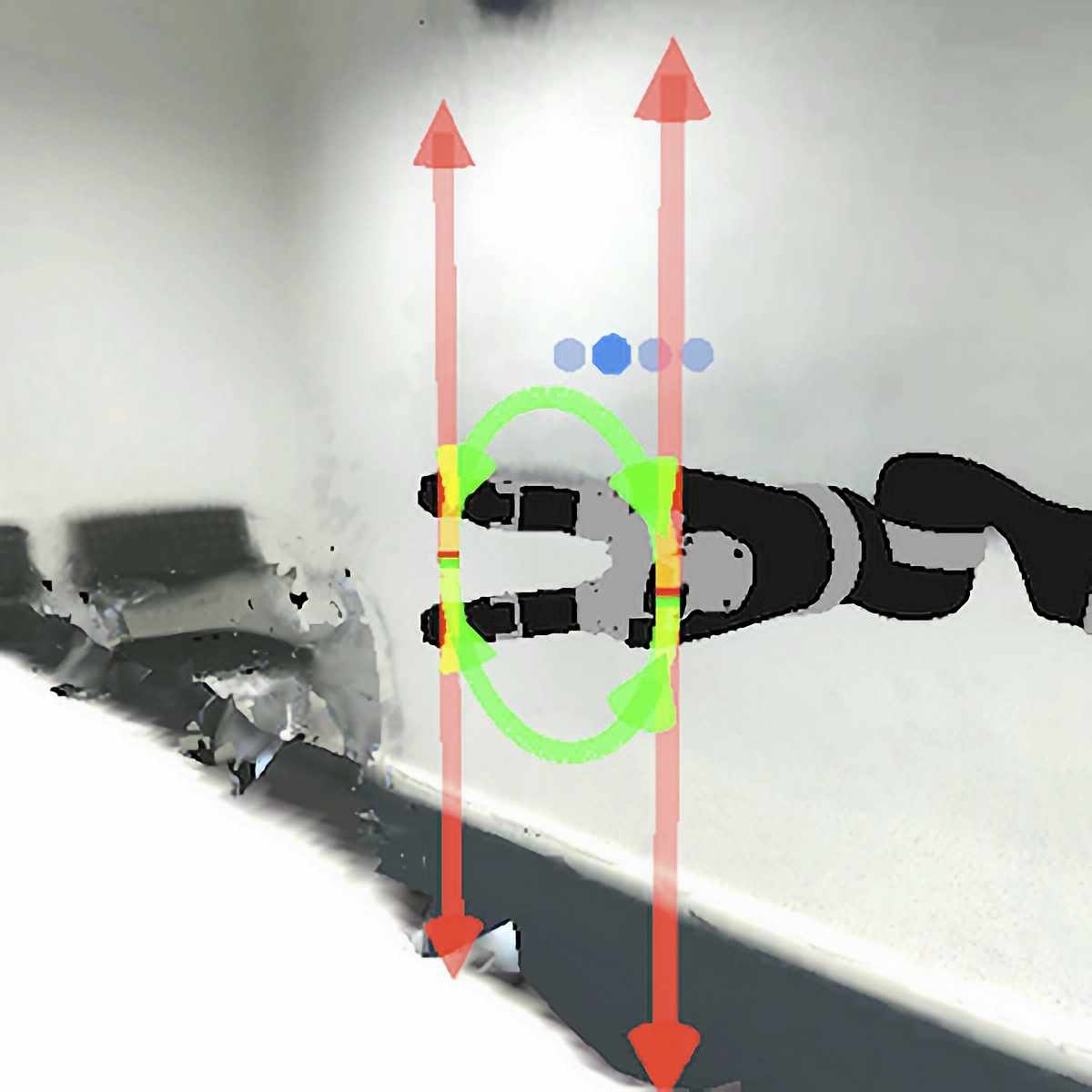}}
  \hfill
  \subfloat[Double Arrow]{\includegraphics[width=0.32\linewidth]{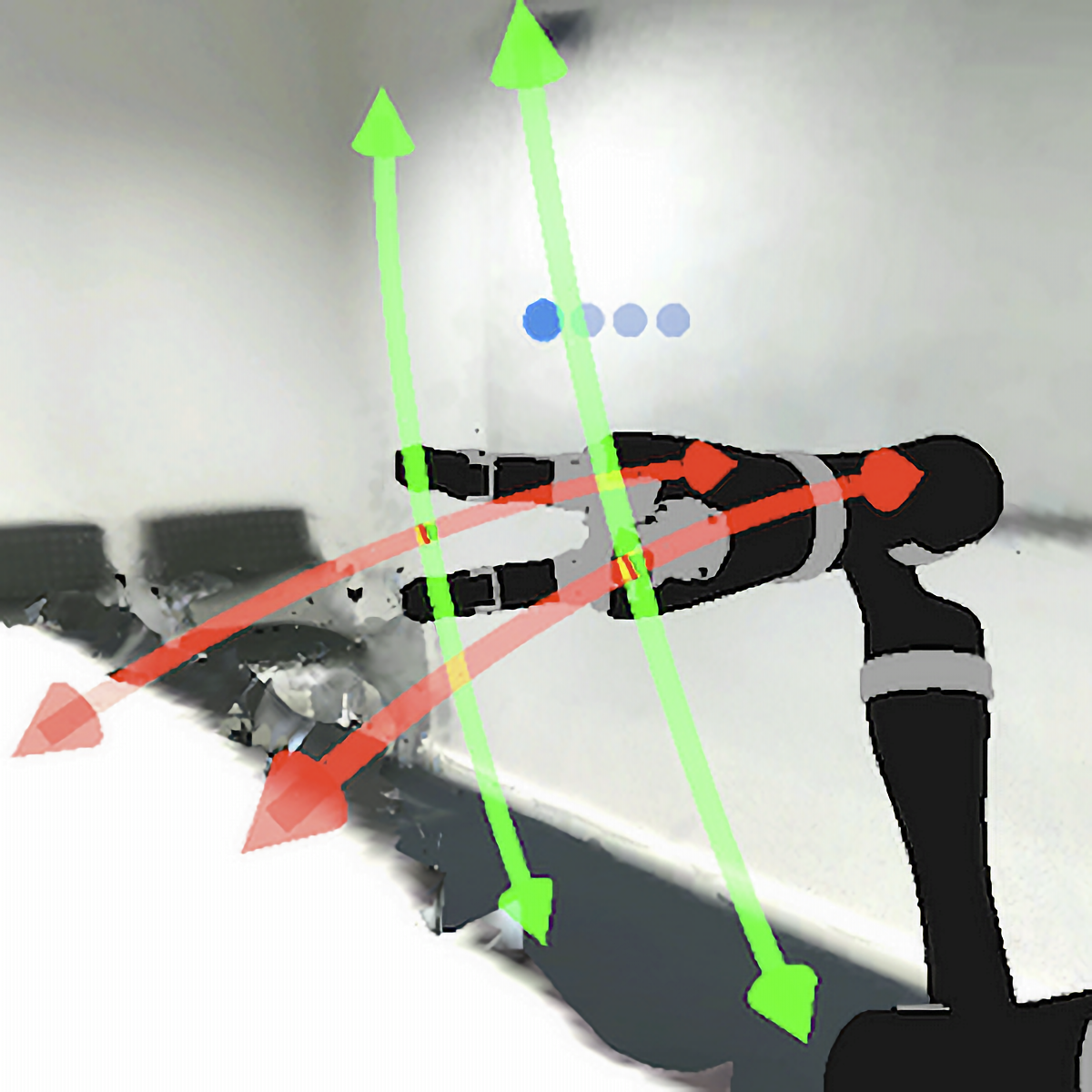}}
  \hfill
  \subfloat[ Single Arrow]{\includegraphics[width=0.32\linewidth]{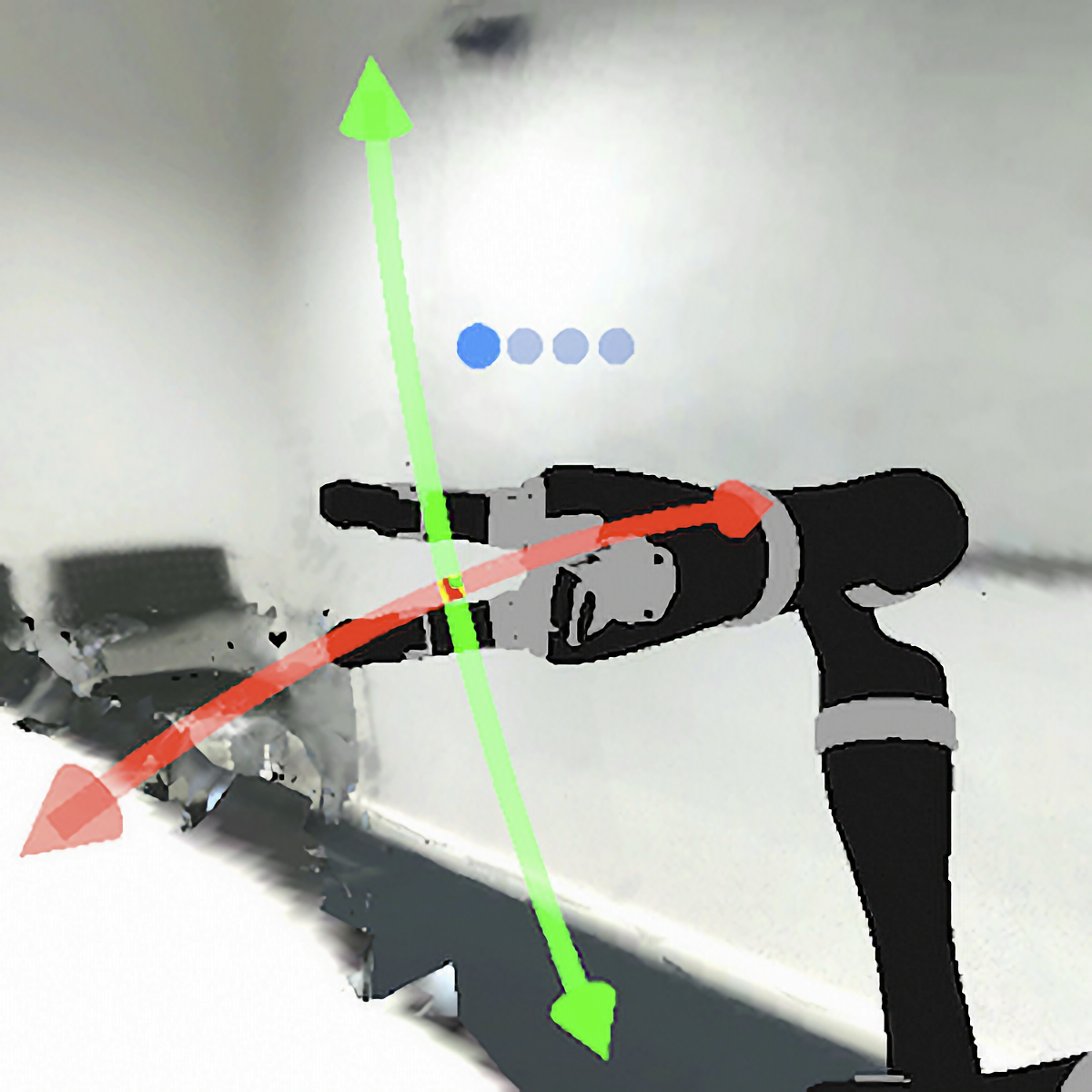}}
\centering
\caption{Evaluated interaction design and visualizations~\cite{Kronhardt.2022adaptOrPerish}.}
\label{fig:visualizations-aop}
\end{figure}

In particular, \change{we} analyzed how the novel adaptive control method -- proposed by \citet{Goldau.2021petra} -- performs in a 3D environment compared to the standard mode-switch approach with cardinal \ac{DoF} mappings. They also investigated whether changes in the visual cues' appearance impact the performance of the adaptive control method.
Three different types of control with varying visual cues and methods of mapping \acp{DoF} were compared in a remote online study. These included the \emph{Classic} visualization, one based on \emph{Double Arrow} using two arrows attached to the gripper's fingers, and a visually reduced variant \emph{Single Arrow}, using only one arrow through the middle of the gripper. See \autoref{fig:visualizations-aop} for a graphical comparison.

Due to the ongoing COVID-19 pandemic, the study was conducted entirely in a \ac{VR} environment created by \emph{AdaptiX}. Non-specific participants were recruited that had access to the required hardware (an \emph{Oculus Quest} \ac{VR}-\ac{HMD}) for an immersive experience. 

The participants repeatedly performed a simple pick-and-place task by controlling the virtual \emph{Kinova Jaco 2} using one of the three control types. Comparative results established that adaptive controls require significantly fewer mode switches than the classic control methods. However, task completion time and workload did not improve. Study participants also mentioned concerns about the dynamically changing mapping of combined \acp{DoF} and the 2-\ac{DoF} input device. 

\textbf{Framework contribution:} \emph{AdaptiX} demonstrated its effectiveness in this remote study to evaluate new interaction designs and feedback techniques. The innovative advantage is that the physical robotic device does not need to be present during these preliminary studies when testing and evaluating essential design elements. The \emph{Record \& Replay} functionality of \emph{AdaptiX} allowed a remote analysis of participants data. This \ac{VR} approach significantly increases the potential to include end-users in the research and design process while at the same time decreasing cost, time involvement, and accessibility concerns.

\subsection{\change{Example Adaption 2}: Communicating Adaptive Control Recommendations}

A follow-up study~\cite{Pascher.2023inTimeAndSpace} evaluated two new adaptive control methods for an assistive robotic arm, one of which involves a multi-modal approach for attention guiding of the user.

\begin{figure}[htbp]
\centering
    \subfloat[]{\includegraphics[width=0.325\linewidth]{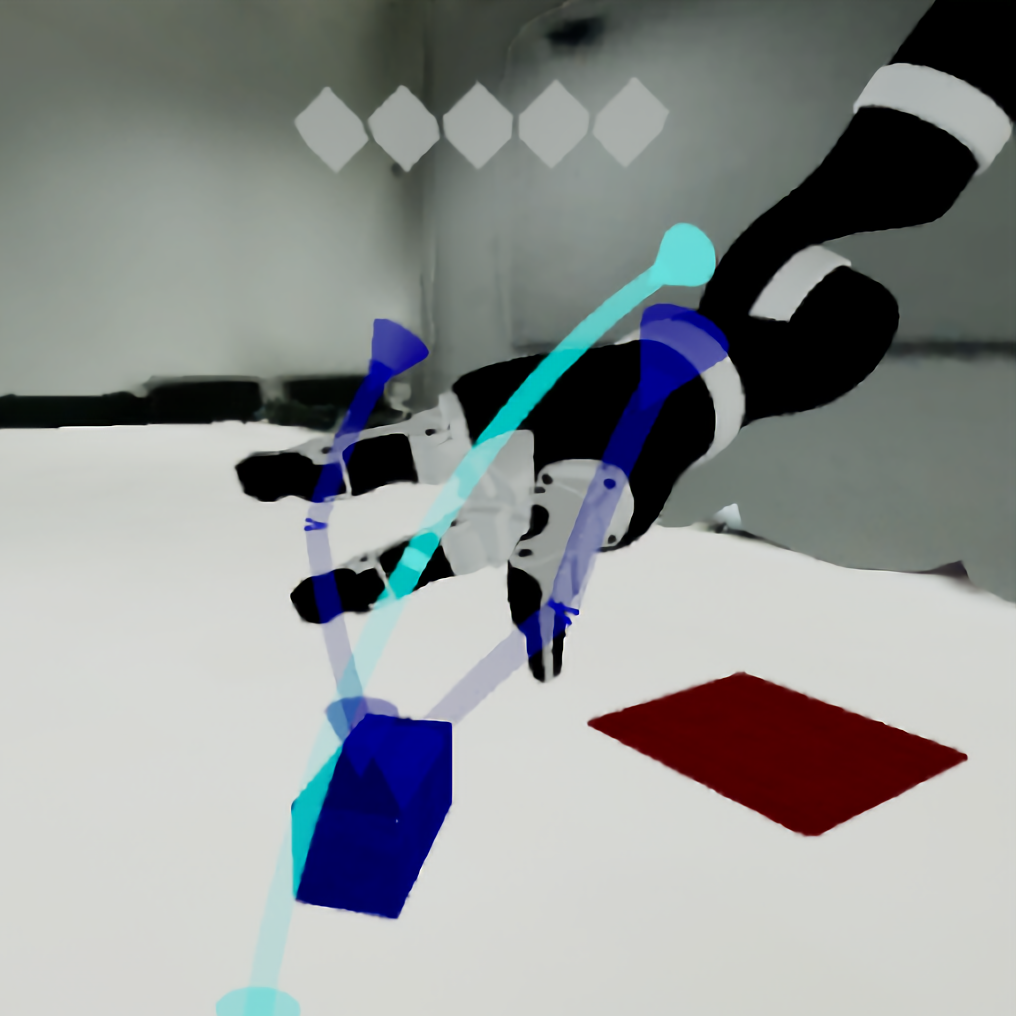}\label{fig:adaptive-suggestions:a2}}
    \hfill
    \subfloat[]{\includegraphics[width=0.325\linewidth]{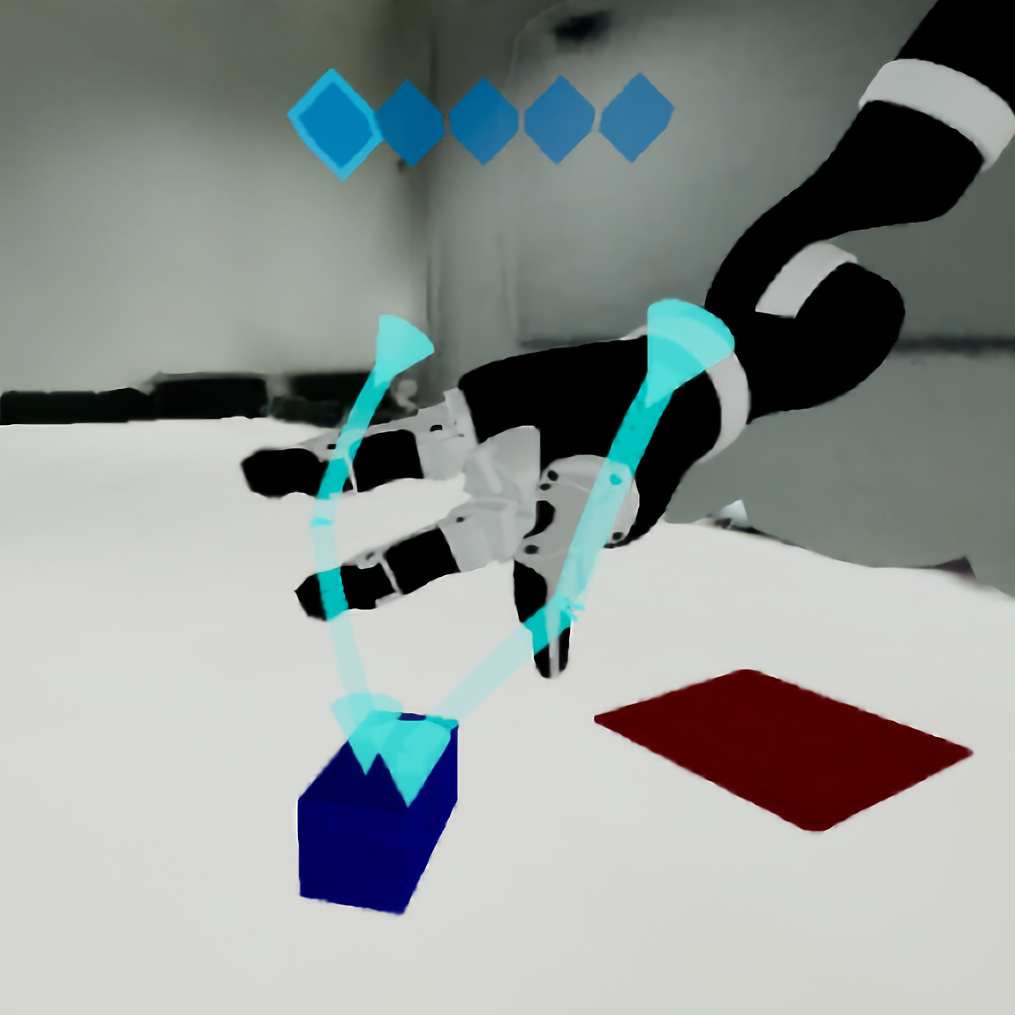}\label{fig:adaptive-suggestions:b2}}
    \hfill
    \subfloat[]{\includegraphics[width=0.325\linewidth]{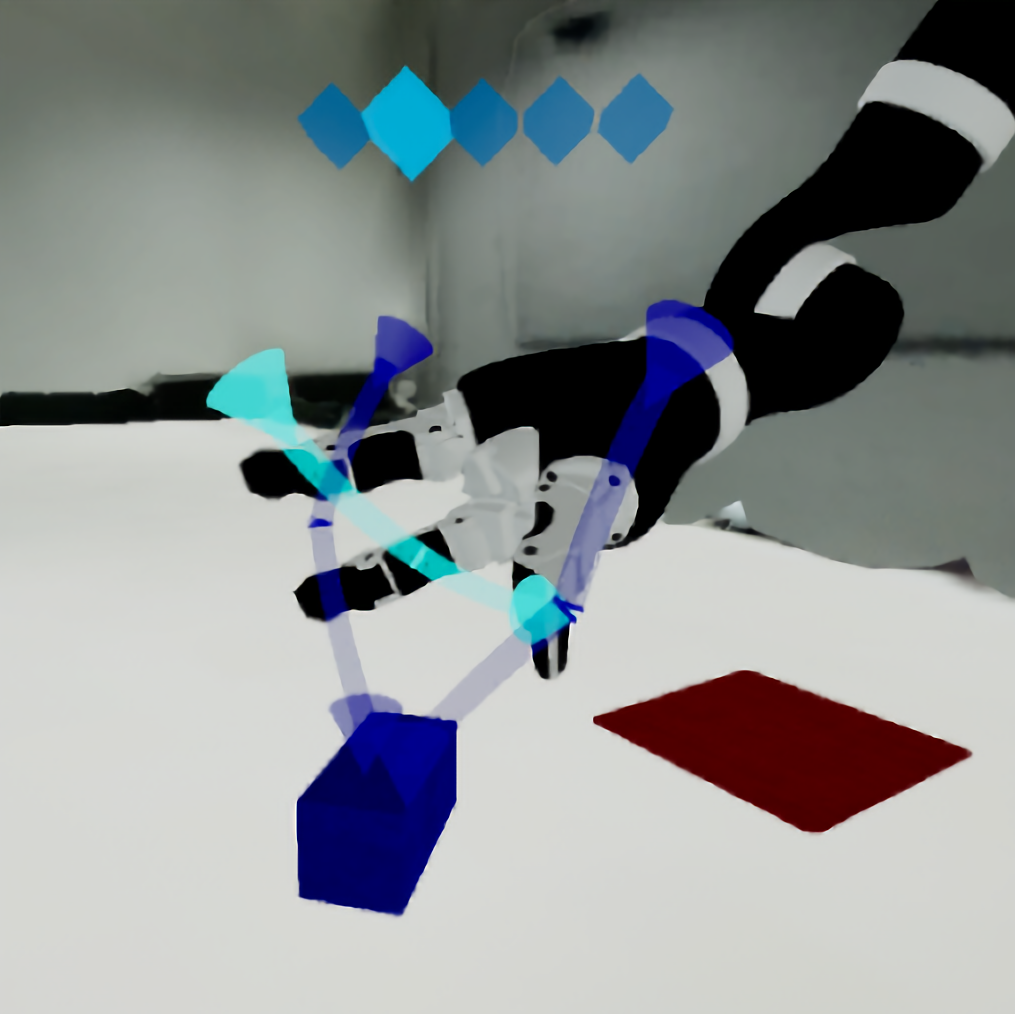}\label{fig:adaptive-suggestions:c2}}
\captionsetup{justification=justified}
  \caption{Suggested control alternatives in light blue, visualized as in case study 2: (\textbf{a}) Moving forward and downward towards the object, (\textbf{b}) Closing the fingers to grasp the object, and (\textbf{c}) Moving towards the target area.}
  \label{fig:visualizations-tas}
\end{figure}

\change{We} used \emph{AdaptiX} in a laboratory study to cross-validate the initial study's findings on how participants interact with the environment. The adaptive system re-calculated the best combination of \acp{DoF} to complete the task during movement. These calculations were presented to the user as alternative control options for the current task. Users cycled through these suggestions -- by pressing a button on the input device -- to make a suitable selection or continue moving with the previous active \acp{DoF} (see \autoref{fig:visualizations-tas}).

They contrasted the variants \emph{Continuous} and \emph{Threshold}, differing in the time at which suggestions are communicated to the user, against a non-adaptive \emph{Classic} control method. Possible effects on task completion time, the number of necessary mode switches, perceived workload, and user opinions on each control method were compared.
Further, we establish that \emph{Continuous} and \emph{Threshold} performed equally well in quantitative and qualitative insights. Consequently, both are promising approaches to communicating proposed directional cues effectively.

\textbf{Framework contribution:} 
The integrated multi-modal feedback is an integral feature of \emph{AdaptiX}, capable of supporting the system's real-time suggestions by user attention guiding. Although some participants experienced the combined visual-auditory-haptic multi-modal feedback as \enquote{irritating}~\cite{Pascher.2023inTimeAndSpace}, it effectively communicated updated suggestions. One application of virtual frameworks like \emph{AdaptiX} might be the differentiation between different modality types and corresponding user preferences in an easy-to-set-up study. 
Highlighting the advantage of our framework, \change{we} could evaluate \change{our} different visualizations and multi-modal feedback without implementing a \ac{VR} environment~\cite{Pascher.2023inTimeAndSpace}.

Based on the successful implementation of \emph{AdaptiX} in this laboratory study, we are confident that the framework performs well in remote and in-person studies.

\subsection{\change{Example Adaption 3}: Comparing Input Devices for Controlling a Physical Robot in Mixed Reality}

A third study~\change{\cite{Pascher.2024inputdevices}} highlights the \ac{MR} capability of the framework and the integration options \change{with} different input devices.
\change{This study used} the \emph{Varjo XR-3} \acs{XR}-\ac{HMD} to explore a similar interaction design and feedback technique to \change{our} \emph{Threshold} approach~\cite{Pascher.2023inTimeAndSpace}. By incorporating this \acs{XR}-\ac{HMD}, the prototype mimics an \ac{AR} environment (see~\autoref{subsec:MRcontinuum}) to the user, seeing the physical setup augmented by visual cues.
Instead of a virtual pick-and-place task as before, this study combined a physical object, a physical drop area, and a physical robotic arm with \ac{AR} cues delivered via the headset. 

Participants compared three assistive input techniques: 1) a \change{\change{head-based control by using the deflection of the head on the \emph{pitch} axis for continuous input and on the \emph{roll} axis for mode-switching}}, 
2) a gamepad input by using the \emph{Xbox Adaptive Controller}~\cite{xboxcontroller} extended with \emph{Logitech Adaptive Gaming Kit}~\cite{logitechkit} buttons for a discrete input, and 3) the control-stick of a \change{\emph{Nintendo Joy-Con}~\cite{joycon}} motion controller -- as a baseline to \change{our previous study}~\cite{Pascher.2023inTimeAndSpace}.

\textbf{Framework contribution:} 
With its real-world setting augmented by virtual cues, the research moved closer to reality on the \ac{MR}-continuum than the previous two case studies. \emph{AdaptiX} successfully performed as an easy-to-use interface between the usage of a physical robot and virtual communication via a \acs{XR}-\ac{HMD}.

It also allowed the research team to quickly evaluate the efficiency of different input devices with the potential to control the robotic arm along the adaptive \ac{DoF} mapping. The standardized \emph{User Input Adapter} enables researchers to easily chose between different technologies -- supporting continuous, discrete, and absolute user input -- and further extend it to their needs \change{by its modular nature}.
\section{Conclusion}
\label{sec:conclusion}
Integrating \emph{AdaptiX} into \ac{HRI} research can streamline the development and evaluation of new interaction designs and feedback techniques for controlling assistive robotic arms. The framework is advantageous in remote and in-person studies as its usage negates the need for a physical robotic device during the initial ideation and prototyping stages, thus increasing flexibility, accessibility, and efficiency.

An initial shared control concept by adaptive \ac{DoF} mapping is provided and implemented to support researchers and developers to either change, extend, or exchange methods with their ideas.
In studies using a physical robot, the integration of \ac{ROS} bridges the gap to reality, by enabling a bidirectional connection between virtual and physical robotic arm. \ac{ROS} allows developers and users to choose between a \emph{DigitalTwin} and \emph{PhysicalTwin} approach while interacting with \emph{AdaptiX}.
Using \emph{AdaptiX}, researchers benefit from the entire continuum of \ac{MR}. As the simulated and real-world environments of the robotic arm are perfectly aligned, nearly seamless switching between controlling the real and virtual robot is possible. This functionality allows applications in pure screen space, \ac{VR}, \ac{AR}, simultaneous simulation/reality, and pure reality.
\emph{AdaptiX}'s 3D teach-in interface facilitates a code-less trajectory programming of an assistive robot by hand-guiding the simulated or real robot to the specific location and saving the position and orientation of the tool center point. These waypoints are interpolated to a combined movement trajectory.
The framework's recording/replaying system is entirely customizable. It includes options to change details during replay, such as repositioning cameras or re-rendering background scenes. A fully integrated recording of participants interacting with the robot is possible, which can be analyzed afterward to evaluate the specific research variables.

Taken together, \emph{AdaptiX} is a free and open-source tool that enables \ac{HRI} researchers to test and evaluate their shared control concepts for assistive robotic devices in a high-resolution virtual environment. The cited case studies clearly demonstrate the benefits researchers and developers can draw from using the framework. The near-endless customization options allow users to tweak the initial version to their specific research needs, resulting in practically tailor-made environments.

\subsection{Framework Extensions}
We invite the community to extend the \emph{AdaptiX} framework based on their requirements needs by creating custom levels/scenarios and integrating new interfaces.
\change{\emph{AdaptiX} can be accessed free-of-charge at \url{https://adaptix.robot-research.de}}.
Refer to the \emph{README} provided in the repository for a detailed description of how to implement experiments in \emph{AdaptiX}.

\begin{acks}
    \change{This research is supported by the \textit{German Federal Ministry of Education and Research} (BMBF, FKZ: \href{https://foerderportal.bund.de/foekat/jsp/SucheAction.do?actionMode=view&fkz=16SV8563}{16SV8563}, \href{https://foerderportal.bund.de/foekat/jsp/SucheAction.do?actionMode=view&fkz=16SV8565}{16SV8565}).} 
\end{acks}

\bibliographystyle{ACM-Reference-Format}
\bibliography{MainPaper}

\appendix
\newpage

\section[Blueprints of ADMC Implementation]{Blueprints of \ac{ADMC} Implementation}
\label{sec:appendix}

\begin{figure}[htbp]
\centering
\includegraphics[angle=90,width=0.45\linewidth]{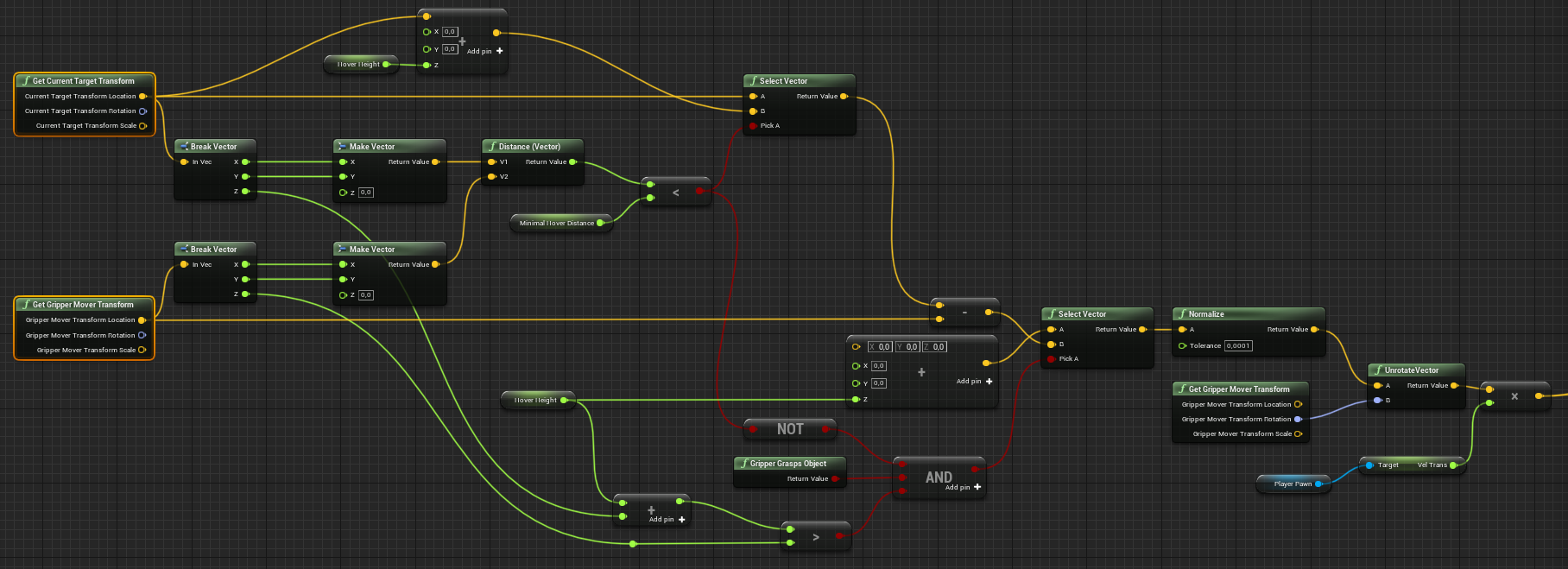}
\centering
\caption{Calculation of the translation for the \emph{Optimal Suggestion}: Excerpt of \emph{Blueprint} code calculating the \emph{Translation} value of the \emph{Adaptive Axis} for the \emph{Optimal Suggestion}. Not pictured: Edge case handling for gripping an object.}
\label{fig:axis-1-translation}
\end{figure}

\begin{figure}[htbp]
\centering
\includegraphics[angle=90,width=0.5\linewidth]{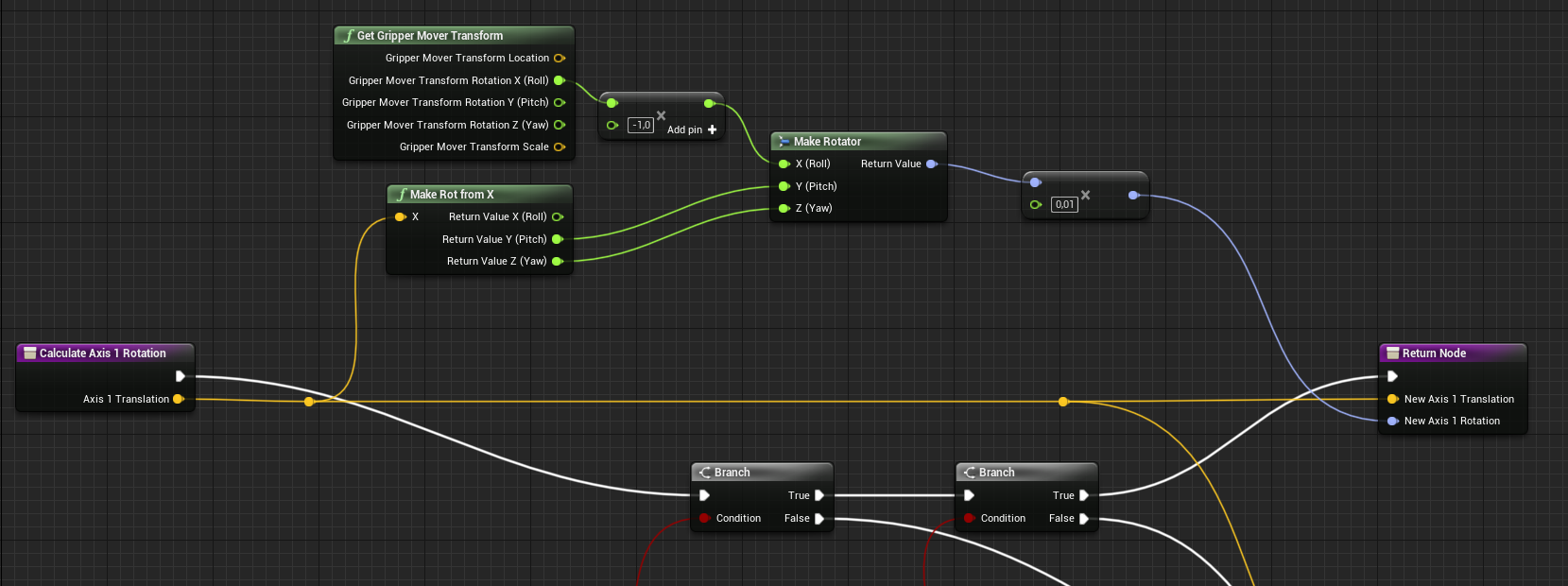}
\centering
\caption{Calculation of the Rotation for the \emph{Optimal Suggestion}: Excerpt of \emph{Blueprint} code calculating the \emph{Rotation} value of the \emph{Adaptive Axis} for the \emph{Optimal Suggestion}. Not pictured: Edge case handling.}
\label{fig:axis-1-rotation}
\end{figure}

\begin{figure}[htbp]
\centering
\includegraphics[angle=90,width=0.6\linewidth]{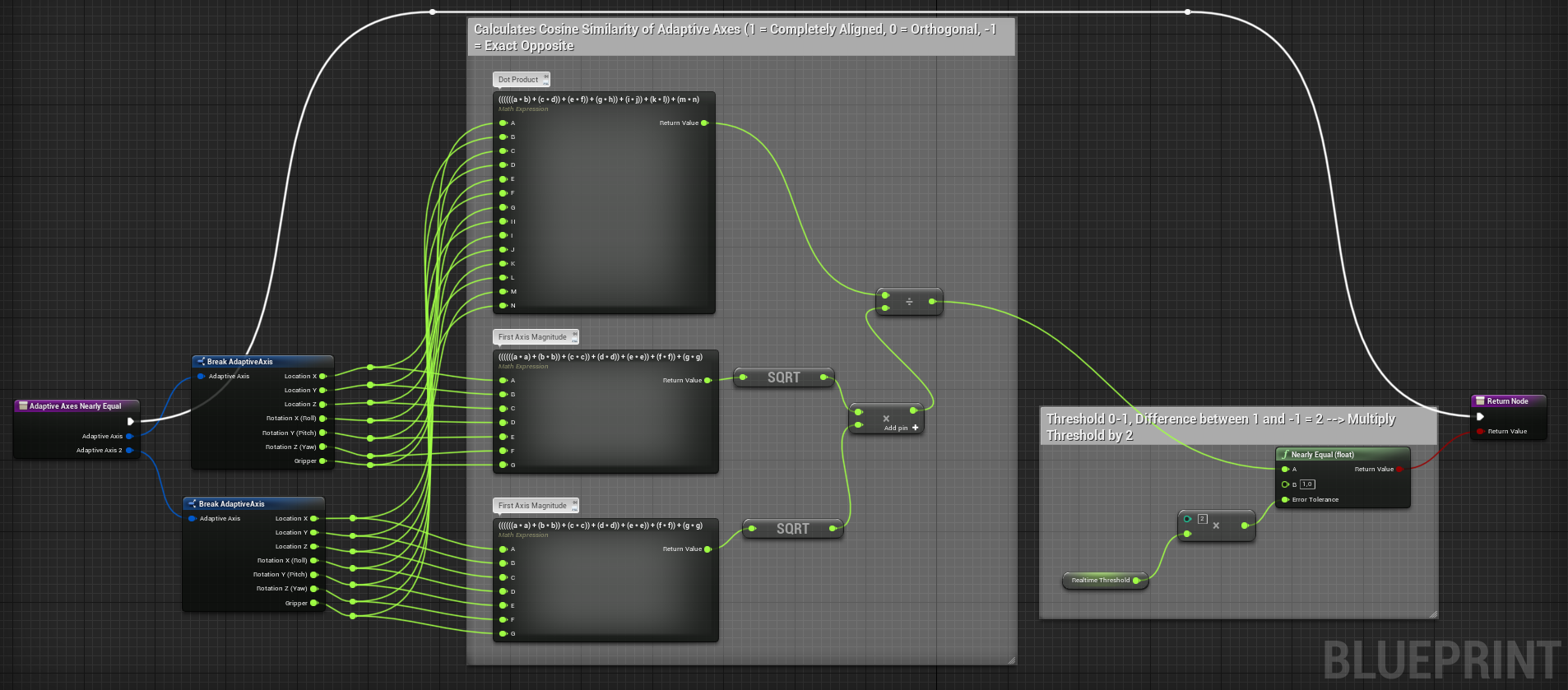}
\centering
\caption{\emph{Adaptive Axes Nearly Equal} function to prepare the multi-modal attention guiding of the user.}
\label{fig:adaptive-axes-nearly-equal}
\end{figure}

\end{document}